\documentclass[twocolumn]{article}
\usepackage{graphicx}

\title{{\bf \Huge Point-Contact Spectroscopy}\\ \normalsize \vspace{1.0cm}
{\it Translation from the russian edition: Moskwa, Izdatelstwo
"Znanije", Serija "Fizika", N12, {\bf (1989)}}}
\vspace{1.2cm}

\author{{\bf \huge Yu. G. Naidyuk, I. K. Yanson}  \normalsize
\\
B. Verkin Institute for Low temperature Physics and Engineering \\
National Academy of Sciences of Ukraine, \\ 61103 Kharkiv,
Ukraine}

\begin{document}

\maketitle

\begin{abstract}
 The history is described of how one of the most commonly used
electric circuit components, an ordinary electric contact, has
become a powerful tool for the physicists to study various
mechanisms of electron scattering in metals. The physical
principles of spectroscopy of quasi-particle excitations in metals
by means of point contacts (PCs) whose dimensions range from only
a few to tens of nanometers are presented in a popular form.

\end{abstract}

\tableofcontents

\section{Introduction}
The most interesting events and phenomena both in everyday life
and science are those which should be characterized by the prefix
'super'. This is usually the case when the objects in question are
under extreme conditions. Thus, on the verge of XX century, a
great impression was made, for example, by the discovery of
all-penetrating invisible X-rays or the phenomenon of resistance
loss (superconductivity) in some metals at low temperatures. Today
the problem of controlled thermonuclear fusion is constantly kept
in the focal point. Its solution, however, involves heating and
confinement of plasma at a superhigh temperature of hundreds
million degrees.

As to the solid state physics, one of the aspects of which is
dealt with in this booklet, the most striking results have been
obtained when studying superpure and perfect or highly disordered
(amorphous), high-quality laminated structures and other compounds
of complex composition. In the first place, the greatest
achievements are gained if the substance under investigation
undergoes the extreme affects of a high magnetic field, low and
super low temperatures, extra high pressure or electromagnetic
radiation.

If all the events described above are associated with the prefix
''super'', how to make them compatible with the word ''point''
used in our case ? It becomes clearer if one deciphers the word
''point'' as super small, i.e. a point contact (PC) stands for a
contact between two substances over a very small area. A question
arises, how small is that area and what it can be compared with.
Naturally, a lower bound is set by the interatomic spacing in a
solid, an upper bound being less clearly defined to within one
micrometer. Since in this case we are interested in the current
passage through such contacts, the term ''super'' can be also
employed to estimate the density of current. Thus the current
densities of 10$^9$ to 10$^{10}$ A/cm$^2$ are achieved in PCs
without substantial heating, though bulk conductors would
momentarily evaporate under these conditions.

The other word, spectroscopy, probably, does not require any
detailed explanation. In a more general sense, spectroscopy is a
method of measuring the energy spectrum of a given object. For
example, when analyzing the emission or absorption spectra of an
atom, one can determine the positions of its energy levels. A
spectrum of different collective excitations also exists in solids
since the atoms are united into a single ensemble. Such
expressions as the energy spectrum of electrons, phonons, magnons,
etc., are well-known in the solid-state physics. Therefore, the
term ''point-contact spectroscopy'' can be interpreted as a method
of measuring the elementary excitation spectra in conductive
solids with the aid of PCs of super small size. In this method,
the contact itself seems to play the role of analytical
instrument, a certain kind of spectrometer.

The point-contact spectroscopy has proved to be a comparatively
simple, available and highly informative method. Within the last
decade, very impressive results have been obtained in this rapidly
developing field by physicists in many countries of the world, as
will be described below.

\section{Contacts Around Us}
In our everyday life, we permanently deal with different electric
contacts by means of which current from the source is transmitted
to the user and distributed over the circuit. Undoubtable, in
number of cases the operation of electrical appliances depends on
reliable contacts connecting different sections of electric
circuits. Consequently, the requirements for trouble free and
stable operation of the contacts are growing from day to day. This
problem has become especially topical due to the rapid growth of
microelectronics when the number of separate elements per unit
volume has increased by a factor of tens of thousands.

From the very inception of electrical engineering, researchers
began to study the electric properties of contacts between
conductors. The results of these activities were summed up by the
German scientist R.Holm in his basic work ''Electric Contacts
Handbook'' published in 1961. Scientists and engineers have
noticed that the electrical resistance of two pieces of metal
tightly pressed to each other is appreciable higher than could be
expected assuming that the electric conductivity at contact is the
same as that inside the conductors. The following reasons have
been advanced. The first one is connected with the fact that any
real surface, even when polished mirror-like, has some microscopic
irregularities, as a result of which in the contact region the
electrodes come in contact only at separate sections. Hence, the
true contact area can be considerably smaller than the apparent
size of contact. The other reason is that the surfaces of metallic
electrodes are always coated with a thin layer of oxides or other
compounds adsorbed by atoms from the air, organic molecules and
other 'dirt'. All these substances, as a rule, possess either
dielectric or semiconductor properties. As a result of it a direct
electric contact is far less than the total area of mechanical
contact between two conductors. Current will not flow in those
places where the dielectric film is sufficiently thick; if the
film thickness does not exceed a few nanometers, part of the
current will pass due to the tunneling effect; metallic
conductivity will be effected only at the part of contact area
which is free from film and 'dirt' because of their damage or
absence. No matter how simple the contact between two electrodes
may look, the real picture can be rather complicated.

The interest in the study of small-size contacts is connected not
only with their technical applications. They proved to be a most
sensitive tool in the hands of scientists. Thus, in the middle of
sixties Academician Yu. V. Sharvin from the Institute of Physical
Problems of the Russian Academy of Sciences proposed to make
experiments in which PCs were used to inject electrons into thin
metallic plates. Then a magnetic field was applied, the injected
beam was focused on the opposite surface of the plate where the
second similar contact was placed to act as a detector. By
changing the magnetic field intensity and relative positions of
contacts on the sample surface, one can observe the focusing of
electrons both along and across the field lines. As to the later
case, such experiments were first made by V.~S.~Tsoi at the
Institute of Solid-State Physics of the Russian Academy of
Sciences. The detector was capable of registering both the
electrons emitted directly by the injector and those once or
repeatedly reflected from the surface. The results of such studies
yielded helpful information on the Fermi surface geometry and
electron scattering by the sample boundary. However, to describe
further investigations more substantially, we will present some
fundamentals of the electronic theory of metals in the next
chapter.

\section{Electrons and Phonons}
An atom is known to possess a definite set of energy levels at
which the electrons reside. When the atoms combine to form a
solid, the wave functions (or orbits for simplicity) of the
valence electrons of neighboring atoms overlap. Thus, the
electrons can move along a crystal, i.e. they actually become
collective and are no longer bound to a definite atom. According
to the quantum picture, the product of particle energy uncertainty
$\delta \varepsilon $ and the lifetime in a particular state
$\delta \tau $ is a constant proportional to the Planck's constant
$\hbar $, i.e. $\delta \varepsilon  \delta \tau \simeq \hbar $.
Therefore, the lifetime $ \delta \tau $ of an electron on a
stationary atomic orbit is large, and thus the energy uncertainty
$ \delta \varepsilon $ is small, i.e. the energy level is narrow
enough. The valence electrons in a solid are no longer bound with
specific atoms, $\delta \tau $ becomes the smaller, the closer are
the ions and the easier it is for the electrons to pass from one
site to another. As a result of it the value of $ \delta
\varepsilon $ increases respectively and the atomic energy levels
become the energy bands $\delta \varepsilon $ wide. If a substance
consists of $N$ atoms, this band will contain $N$ energy levels at
which all the collective electrons reside.

\begin{figure}[t]
\includegraphics[width=8.5cm,angle=0]{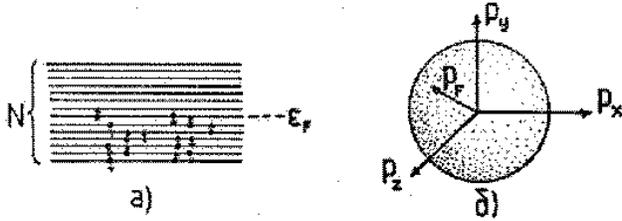}
\caption[]{\small a) Electron distribution over the energy levels
in a solid consisting of N atoms on condition that each atom gives
away one electron into the conduction band. According to the Pauli
principle, each level can accommodate two electrons with opposite
spins, then the conduction band will be half-full and Fermi level
separating the filled states from the free ones passes in the
middle of the band.  b) The Fermi surface for free electron gas in
a solid in the momentum space. At $T=0$ it is an isoenergetic
surface  separating the filled electron states from the free ones.
In real metals the Fermi surface is close to a sphere only in some
alkali metals, in the other it has more  complicated topology and
can consist of several parts. The geometrical characteristics of
the Fermi surface, its  shape, curvature, extreme crossections
values determine the main electronic properties of metals, thus
enabling  reconstruction of its form from the experimental data. }
\label{fig1}
\end{figure}
\normalsize

In the quantum mechanics, the microparticle behavior depends on
whether they have an integer (0, 1, 2, ...) or half-integer (1/2,
3/2, 5/2, ...) spin. A spin is a quantum characteristics of a
particle which is associated with its intrinsic angular momentum.
The number of particles being in the same state with the same
energy is unlimited for those with an integer spin named bosons.
On the contrary, only one particle with a half-integer spin, a
fermion, can exist in a state with similar parameters. Electrons
are fermions with a 1/2 spin, as a result of which only two
electrons with opposite spins can be located at the same energy
level. According to this principle, at a temperature close to
absolute zero all the electrons cannot reside at the minimum
energy level and are forced to successively fill the higher-lying
energy levels in the allowed band. Coming back to the band in a
monovalent solid consisting of $N$ atoms comprising $N$ states, it
becomes clear that this band can hold $N$ collective electrons
starting from the minimum energy level (band bottom) and up to the
half-height (Fig.\,\ref{fig1}). The highest level occupied by
electrons at $T=0$ was called the Fermi level, and the energy of
this level - the Fermi energy $ \varepsilon_F$ which corresponds
to the Fermi momentum \footnote {It should be noted that the
electron behavior in a lattice is quite different from that in a
vacuum. In fact, one cannot consider a separate electron in a
solid, but rather refer to electron-like excitations, i.e.
quasi-particles with a charge similar to that of a free electron
and a close mass. This is why the dispersion law includes the
effective mass value. Further on we shall use the term 'electrons'
implying thesequasi-particles.} $p_{F}=(2m_{el}
\varepsilon_{F})^{1/2}$. In the momentum space, the Fermi level is
an isoenergetic surface (Fermi-surface) which separates the
occupied states from the vacant ones. For the case of the
isotropic dispersion law, when the energy depends only on the
momentum magnitude, but not on its direction, the Fermi surface is
a sphere with the radius $p_F$ (Fig.\,\ref{fig1}(b)).

The electrons can move within the lattice, the number of particles
moving, for example, from right to left, being exactly equal to
the number of those moving in the opposite direction, while the
state of the system remains unchanged due to the
quantum-mechanical indistinguishability of the particles. To
create current in the crystal, it is necessary to transfer part of
electrons into a state in which they can move under the action of
the field. Since the number of atoms in a solid is about $10^{22}$
per 1 cm$^3$ and the Fermi energy is 1 to 10 $eV$ or $10^4$ to
$10^5 $K, the distance between the nearby levels proves to be very
small, equal to $\varepsilon_N \simeq 10^{-21}$ to $10^{-22}$ eV
or $10^{-17}$ to $10^{-18}$ K. Thus, the electrons residing at the
Fermi level under the action of the field move to the nearest free
higher lying level separated by an extremely small energy gap of
about $10^{-18}$ K, i.e. can easily increase their energy or
accelerate. Such substances with unfilled electron band readily
conduct the electric current and are metals. As to the
dielectrics, their bands are filled and, as a rule, the nearest
vacant band is separated by an energy gap of a few electron-volts.
This is possible when two electrons of each atom enter the band.
Now the electrons cannot change their states under the action of
the field as in this case they have to "jump over" a large energy
gap and there will be no current. Note that the contribution into
conduction is made only by the electrons residing near the Fermi
level while the lower-lying ones cannot change their states since
the higher lying levels are occupied. Therefore, all the kinetic
properties of conductors are mainly determined by a small group of
electrons residing near the Fermi level in a layer with a width
about the thermal smearing $k_BT$.

Now we will try to obtain an expression for calculating the
conductivity of metals. The current density $j$ can be written as
a product of average velocity of ordered electrons motion $\langle
v\rangle$, their density $n$ and charge $e$: $j=ne\langle
v\rangle$. In a field with the intensity $E$ an electron acquires
acceleration $a=F/m=eE/m$, whence $\langle v\rangle$ can be
estimated as $\langle v\rangle \simeq a\tau = eE \tau/m$, where
$\tau$ is the mean free time of electrons between their scattering
by impurities, defects, phonons, etc. As a result, the current
density can be written as $j=ne^2 \tau E/m=E/\rho$, where
$\rho=m/ne^2\tau$ is the resistivity of metal. It can be seen that
the current in metals is proportional to the applied field, i.e.
the well-known Ohm's law holds. Let us rewrite the expression for
$\rho$ in the form

\begin{equation}
\rho = m/ne^2 \tau = mv_F/ne^2 \tau v_F = p_F/ne^2 l \label{rhol}
\end{equation}
where $v_F$ is the Fermi velocity, $l=v_F \tau$ is the mean free
path (mfp) of electrons. It evidently follows from Eq.
(\ref{rhol}) that the less is the mfp of electrons in the lattice,
the higher is its resistance.

On which factors does the mfp depend? It has been already
mentioned that the electrons are scattered by the impurity atoms,
defects, sample surface, etc. At first glance, it appears that
electrons should rather "stumble" on ions in the lattice. However,
it turns out that according to the laws of quantum mechanics the
electrons are scattered only by thermal vibrations of atoms - the
phonons. In a strictly periodical perfect lattice at $T=0$ the
electrons move without collisions as if they do not notice the
ionic framework. This is another consequence of the fact the
properties of conduction electrons in a solid differ but radically
from those of free electrons in the vacuum.

The probability of electrons scattering is proportional to the
frequency of their collisions, i.e. to the inverse relaxation time
$\tau^{-1}$. Considering all the scattering mechanisms as
independent ones, the total probability of scattering under the
actions of all factors will be equal to the sum of separate
probabilities $ \tau^{-1}  =  \tau^{-1}_i + \tau^{-1}_p +
\tau^{-1}_e + \ldots $, or, which is the same, $ \rho  =  \rho_i +
\rho_p + \rho_e  + \ldots $  as $ \tau^{-1} \sim \rho $. Here $
\rho_i $, $ \rho_p $ and $ \rho_e $ are resistivity associated
with the scattering by impurities and defects (i), by phonons (p),
by electrons (e). We have obtained the so-called Matthiessen rule
according to which the total electrical resistivity is the sum of
contributions due to separate scattering mechanisms.

We will be more interested in the term $\rho_p$ determining the
resistivity due to the electron-phonon interaction. First of all,
we must make clear what a phonon is. The vibrations of lattice
atoms can be presented in the form of excitation waves propagating
in a crystal. It is appropriate here to draw on the analogy with
waves on the surface of water. A float immersed in the water will
move up and down during the wave propagation, i.e. it will
vibrate. Thus, the vibrations of atoms in a solid are described in
the language of waves. According to the quantum mechanical
dualism, each wave can be compared to a particle and the thermal
vibrations of lattice atoms can be described in terms of
quasiparticles - the phonons. The wave vector of the wave $k$
corresponds to the momentum of particle  \footnote{ Since the
wavelength $\lambda$ of atom vibrations in the lattice cannot be
smaller than the interatomic distance $a$, namely $\lambda_{min} =
2a $, the maximum momentum of phonons is $q_{max} = \hbar k_{max}
= \hbar 2\pi/\lambda_{min} =\hbar \pi/a$. For metals $q_{max} \sim
p_{F}$.}$q=\hbar k$. In such a model the electrons scattering by
thermal vibrations of the lattice can be considered as the
interaction of quasi-particles, electrons and phonons, making use
of the familiar laws of conservation of energy and momentum.

\begin{figure}[t]
\includegraphics[width=8cm,angle=0]{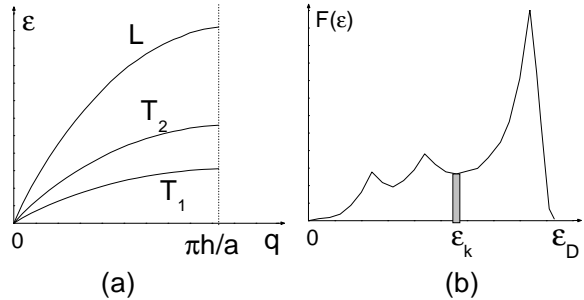}
\vspace{-1.5cm}
  \caption[]{\small a) Phonon dispersion law in metals with one
atom per unit cell. Two lower curves correspond to the transverse
vibration of atoms with two polarization, the upper curve
corresponds to the longitudinal vibrations, compression-extension
waves. b) Typical phonon spectrum for a solid. The boundary Debye
energy $\varepsilon_{D}$ is determined by the maximum  possible
frequency of atom vibrations in the lattice. As a rule, the maxima
associated with low group velocity phonons $v\sim d\varepsilon/dq
\rightarrow 0$. In Fig.\,\ref{fig2}(a) it corresponds to the
phonons with maximum momentum  $q_{max}=\pi\hbar/a$.} \label{fig2}
\end{figure}
\normalsize

The energy of phonon $\varepsilon_{ph}$, as well as that of
electron, depends on its momentum. The dispersion law for phonons,
i.e. the $\varepsilon_{ph}$ dependence on $q$, is more complicated
than for electrons for which the formula $\varepsilon =
p^2/2m_{eff}$ gives a fairly good approximation. It is not easy to
describe this law analytically over the whole range from 0 to
$q_{max}$ and it is usually obtained experimentally. Besides, the
phonon dispersion law depends not only on the value of $q$ but
also on its direction. Fig.\,\ref{fig2}(a) shows the dispersion
curves for metals with one atom in the unit cell and one direction
in a crystal. A more convenient and more often used
characteristics of phonons in a solid is the so-called phonon
spectrum $F( \varepsilon)$. This dependence reflects the phonon
density of states or, in more detail, indicates which part of
phonons has the particular energy. Fig.\,\ref{fig2}(b) shows the
typical form of the $F(\varepsilon)$ curve in a crystal. The first
two maxima are generally due to transverse vibrations of atoms in
two perpendicular directions (polarizations) while the third
maximum is due to more energized longitudinal vibrations since
their excitation in a crystal causes the compression and extension
waves. As is seen from Fig.\,\ref{fig2}(a) there exists a maximum
energy of phonons, or in the wave language, the boundary (Debye)
frequency of vibrations $\omega_{max} (\hbar \omega_{max} =
\varepsilon_{max}$). The Debye frequency expressed in degrees is
close to so-called Debye temperature. It has a very simple
physical meaning: This is a temperature at which the whole
vibrational spectrum of metal, up to the highest-energy phonons,
is excited.

The phonon spectrum is an important characteristic of a solid. It
enables determination of thermal characteristics of the lattice,
the heat capacity and heat conductivity, plays a decisive role in
kinetic phenomena stipulated by the electron-phonon interaction.

\section{Waste recovery}
Such a heading would be most suitable for popular-scientific
literature on economy or ecology. It occurs, however, that
sometimes the "waste" can be utilized in scientific research as
well. Turning closer to the point, we will first describe one of
wonderful phenomena - the tunneling effect. One of the paradoxical
predictions of quantum mechanics which appeared at the beginning
of this century was the possibility of microparticle penetration
through a potential barrier. In the common language it would mean
that, for example, a ball strikes against a wall and passes
through it without causing any injuries. This phenomenon was
called tunneling, naturally bearing in mind that no hole is left
after the particle passage. The probability of tunneling is the
higher, the smaller are the thickness and height of potential
barrier (wall thickness and hardness) to be overcome. The
tunneling effect was immediately used to explain such microcosm
phenomenon as $\alpha $-decay of radioactive nuclei. However, the
most striking manifestation of tunneling could be observed in
solid-state physics of the last sixties.

Let us see what will occur when a potential difference is applied
to two metallic plates arranged like in a capacitor and then the
interplate distance is gradually decreased. Let the plate
potential be not very high to avoid electric breakdown or suppose
they are in vacuum. It turns out that at a certain plate
separation when they are not yet in contact, current appears due
to electrons tunneling through the gap, which in this case serves
as a potential barrier. The interplate distance at which these
effects arise proves to be very small, not greater than a few
nanometers ($10^{-9}$m). It is rather difficult to create and
stationarily maintain such a vacuum gap between conductors
\footnote {It should be noted that in the beginning of eighties it
became possible to make precision instruments where tunneling
proceeds through a vacuum or air gap between a flat surface and a
sharp needle. As the tunneling current strongly depends on the
amount of gap, by moving the needle over the surface and
simultaneously approaching it to or withdrawing it from the sample
at constant tunneling current, one can take surface relief with a
superhigh resolution (not worse than $10^{-10}$m). Such
instruments were called tunneling scanning microscopes.} ,
therefore physicists first took a different line. The gap between
two evaporated films of metal was formed from dielectric oxides,
one being oxidized before evaporating the other. The oxide layer
on a metal, as a rule, is few nanometers thick and serves as a
good and strong dielectric. This structure was called a sandwich
or a tunneling contact. The latter proved to be interesting not
only due to the observation of the tunneling effect, but also as a
tool of physical research.

\begin{figure}[t]
\includegraphics[width=6.5cm,angle=0]{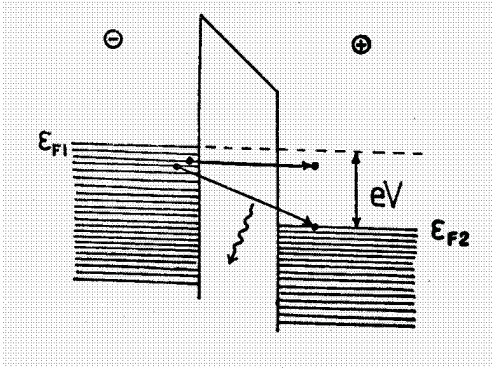}
\caption[]{\small Energy scheme of the tunnelling contact formed
between two  metals separated by a thin layer of dielectric to
which the potential difference V is applied. Two possible channels
of electron tunnelling are show: the elastic one with energy
conservation and inelastic with energy loss and phonons or
molecules excitation in the barrier or on its surface. }
\label{fig3}
\end{figure}
\normalsize

When two metal come in contact, their Fermi levels coincide for
otherwise the electrons would transfer from the metal with a
higher energy to the metal with a lower one. If a voltage $V$
(Fig.\,\ref{fig3}) is applied to the tunneling contact, the Fermi
level will shift with respect to one another by the value $eV$.
The electrons residing on the left can tunnel into the right metal
as there are free states on the right. The Ohm's law holds in such
a contact even at not very high current voltages. The situation
slightly differs if in the process of tunneling the electron
changes its energy after the interaction with, for example,
molecules absorbed on the dielectric layer surface, phonons in the
bulk metal or in oxides separating the metals. After the energy
loss an electron tunnels to other states, i.e. it obtains a wide
choice of places to get to. Such processes will result in the
tunneling current increase and deviations from the Ohm's law. Such
correction is small and for its exact observation one can usually
plots not current-voltage characteristics but their derivatives
which make it possible to determine the energy of quasi-particles
or excited molecules interacting with electrons. This method of
study was called tunneling spectroscopy and has been widely used
in solid-state physics. The most complicated thing in this
technique is to create a thin dielectric layer between the metals.
If the oxidation is strong, the layer will be thick and the
tunneling current negligibly small. If one tries to reduce the
oxidation time, the appearance of 'holes' becomes possible, i.e.
the dielectric film will be discontinuous and conductivity will be
stipulated not by tunneling but by current flowing throw the
metallic shorts. Such contacts are naturally to be rejected for
they are no good for tunneling investigations.

\begin{figure}[t]
\includegraphics[width=7cm,angle=0]{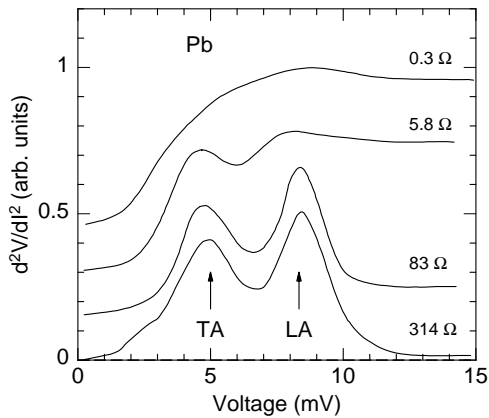}
\caption[]{\small Second derivatives of current-voltage
characteristics of a  shorted out tunnelling contact of lead
recorded with its resistance decrease from 314  to  0.3 Ohm.  The
spectral maxima correspond to those of the phonon  density of
states in Pb marked by arrows.} \label{fig4}
\end{figure}

One of the authors (I. K. Yanson) has been engaged with the
problems of tunneling spectroscopy for a long period of time. At a
certain stage he paid attention to this rejected tunneling
contacts with metallic shorts "wasted" from the tunneling
measurements. It turned out that the current-voltage
characteristics of shorted-out contacts also displayed deviations
from the Ohm's law. In contrast to the tunneling contacts where
current increased more rapidly than linearly, i.e. the contact
resistance decreased with the increase of the voltage, in
shorted-out samples the resistance increased. One day, while
investigated a shorted-out tunneling contact made of lead, the
scientist found out that the second derivative of its
current-voltage characteristics looked amazingly similar to the
phonon density of states in this metal (Fig.\,\ref{fig4}). It
could not happen accidentally since when shorting out this contact
by current pulses, i.e. increasing the constriction dimensions,
the picture did not change and the maxima retained their
positions. Later on this was checked on shorted-out contacts from
other metals: copper, silver, gold, tin, etc. Similar pictures
were observed in all cases : the second derivative of the
current-voltage characteristics reminded of the phonon density of
states in metals under investigation.

\normalsize

\section{What happens inside the contact }
The question arises why the above-mentioned shorts revealed such
spectroscopic properties, what kind they are and what size they
have. One may suppose that in shorted out tunneling contacts the
size of 'holes' is comparable to the dielectric thickness, that is
a few nanometers. But the size of a metallic constriction can be
determined more precisely by the its resistance.

\begin{figure}[t]
\includegraphics[width=8.5cm,angle=0]{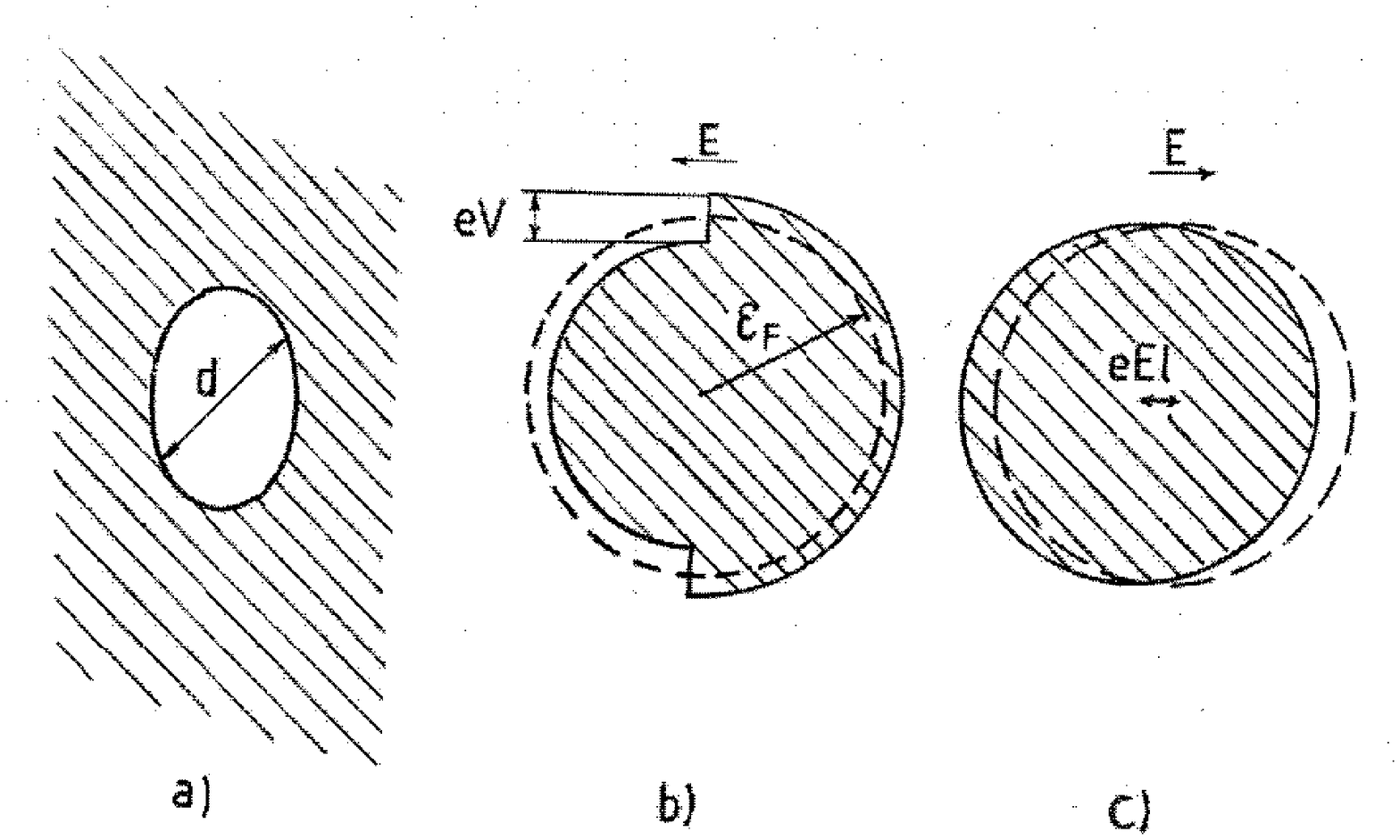}
\caption[]{\small a) Model of PC in the form of a thin
electron-non-transparent screen separating two metallic
half-spaces with the orifice diameter $d$.  b) Central section of
the Fermi surface in the center of  PC to which the voltage $V$ is
applied.  c) The same in a homogeneous conductor with the electric
field intensity E. here under the field effect the whole fermi
surface is shifted by eEl (l is the electron mfp) which is much
smaller then eV.} \label{fig5}
\end{figure}
\normalsize

Consider  PC model in the form of a thin, electron-opaque
partition with a hole $d$ (Fig.\,\ref{fig5}(a)). The expression
for the resistance of such contact proved to be dependent on the
ratio of its size $d$ and the electron mfp $l$. The case of $l \ll
d$ was considered by the famous physicist J. Maxwell already at
the end of the XIX century. The resistance of such contacts with
large diameters or contacts formed by metals with short electron
mfp is expressed rather simply as $R_M = \rho/d$. In the case of
the reverse ratio, when $l \gg d$, the so-called ballistic regime
of the electrons flight is realized, i.e. they pass through an
orifice along rectilinear trajectories. In this case, as was shown
by Yu.~V.~Sharvin in the middle of sixties, $ R =16\rho l/ 3\pi
d^2$. The expression for contact resistance at random $l/d$ ratio
was obtained in 1966 by G.~Wexler
\begin{equation}
R = \Gamma \rho/d + 16\rho l/3\pi d^2 \label{Rwex}
\end{equation}
Here the value of $\Gamma$ varies from
0.7 to 1 with the decrease of $l/d$ ratio from $\infty $ to zero.
It can be seen that the Wexler formula is actually the algebraic sum of two
resistances, those of Maxwell and Sharvin.

Let us evaluate the size of lead contact mentioned in the previous
chapter. The contact resistance was 314 $\Omega$ and from Eq.
(\ref{Rwex}) we will find that its diameter is about 2 nm or 20
\AA , i.e. it is really close to the thickness of oxide \footnote
{When calculating, the value of $\rho l$ was set equal to
$10^{-11} \Omega $cm$^2$ and $\rho \sim 10^{-6} \Omega $cm. These
values can be used for most metals.}.

We have already considered the factors responsible for the mfp
$l$, but to compare it with the contact size one should find its
absolute value. In metals at low temperatures, in the absence of
phonons and other quasi-particle excitations, $l$ is determined by
scattering due to various lattice defects : impurities,
dislocations, sample boundaries. In extremely pure and perfect
single-crystalline samples at helium temperatures mfp may reach
fantastic values of tens and hundreds thousands of interatomic
distances, i.e. be measured in millimeters. With the increase of
temperature and phonons excitation, the electron mfp decreases due
to scattering by phonons. At this time $l$ can reduce to tens of
nanometers.

It was noticed in the initial experiments on the investigation of
PCs that the spectrum was most distinctly visible for contacts
with higher resistance whose size was of a few nanometers,
respectively. At the same time, the spectral features smeared and
gradually disappeared in the case of large-size low-resistance
contacts (Fig.\,\ref{fig4}). It become clear that distinct
features were observed in samples for which the electron mfp was
greater than the contact size, or when the electrons flew through
contacts along the ballistic trajectories.

If the electrons flow through the contact without scattering, the
contact resistance is expressed by the Sharvin formula, the
current-voltage characteristics will be linear, its second
derivative will be equal to zero and, naturally, there will be no
spectrum. The situation changes if we take into account that part
of electrons are scattered by phonons in the contact neighborhood.
According to the calculations, the probability of this scattering
is proportional to $ d/l_{e-ph}$, where $l_{e-ph}$ is the length
of electron-phonon interaction. At low temperatures, when the
number of phonons is small, the processes of excitation
(generation) of phonons by electrons near the constriction
predominate. As a result of it part of electrons will not get into
the hole and the total current will decrease. This means that the
contact resistance will rise due to the electron-phonon
interaction, thus causing deviations from the Ohm's law and the
signal appearance in the second derivative.

Qualitative considerations are also appropriate here. At the
increase of voltage by $\Delta V$, the change of resistance
$\Delta R $ will be proportional to the probability
$G(\varepsilon)$ of excitation by electrons in the contact zone of
a phonon with the energy $\varepsilon$, i.e. $\Delta R \propto
G(\varepsilon) \Delta V $ or, taking into account that $R \propto
\Delta V / \Delta I$, we shall evaluate the second derivative as
$d^2V/dI^2 \propto \Delta R/ \Delta V \propto G(\varepsilon)$.
Thus, the second derivative of the current-voltage characteristics
reflects the probability of phonons emission $G(\varepsilon)$. The
latter, in its turn, is proportional to the phonon density of
states $F(\varepsilon)$ multiplied by the so-called square of the
electron-phonon interaction $\alpha_{PC}^2 (\varepsilon)$ taking
into account the "force" of electrons interaction with one or
another group of phonons and the contact geometry. Therefore we
can write
\begin{equation}
d^2V/dI^2 \propto G(\varepsilon) = \alpha_{PC}^2(\varepsilon)
F(\varepsilon)
\end{equation}
This expression was later rigorously substantiated in theoretical
works described below. It turned out that the
$\alpha^2(\varepsilon)$ dependence is "smoother" function of
energy compared with $F(\varepsilon)$, so the shape and
peculiarities (maxima, minima) of $d^2V/dI^2$ are to a greater
extent dictated by the phonon spectrum behavior. Now it is clear
why the spectra for lead and other metals remind us the phonon
density of states.

\section{Some theoretical fundamentals }
The theory of point-contact spectroscopy has been developed by I.
O. Kulik and his coworkers A. N. Omelyanchuok and R. I. Shekhter.
Their principal result was rigorous derivation of the expression
for $d^2V/dI^2$ of PC in the ballistic regime and its relation to
the function $F(\varepsilon)$. A model chosen for PC was an
orifice in a thin opaque to electrons partition separating two
metallic half-spaces (Fig.\,\ref{fig5}(a)). Later on it has been
demonstrated that the result is noncritical with respect to the
contact geometry. The expression obtained theoretically has the
form
\begin{equation}
\frac{d^2V}{dI^2} \simeq - \frac{d^2I}{dV^2} \propto \langle
K\rangle^{-1}d^3 N(\varepsilon_F) G(\varepsilon) \label{aF}
\end{equation}
where $\langle K\rangle$ is the PC geometry factor equal to 0.25
for the orifice model, $d$ is the contact diameter,
$N(\varepsilon_F)$ is the electron density of states at the Fermi
surface.

The nonequilibrium function of electrons distribution in the
center of contact is shown in Fig.\,\ref{fig5}(b). When voltage is
applied to the contact, the electrons are separated into two
groups arriving from the right and the left regions,
correspondingly. Maximum energies of these groups at $T=0$ differ
precisely by the amount of $eV$. It becomes important that the
energy difference between any two states of electrons at the Fermi
surface is equal to $eV$ or zero. It is this circumstance that
enables us to use PCs as an unusual energy "probe" to determine
the energies of other quasi-particles.

Fig.\,\ref{fig5} shows that the electrons distribution function in
PC differs but greatly from a similar one in the case of current
flowing through a homogeneous sample. In this case it is
impossible to impart a considerable excess energy to the electrons
that is comparable to the energy of Debye phonons. It turns out
that for the electron energy acquired in the electric field to
reach the characteristic phonon energies, the current energy
should exceed $10^9$ A/cm$^2$. Such current densities in the bulk
metal are unattainable because of the Joule heating and melting.
It is not the case with PCs as due to their small size the heat is
quickly 'sucked off' into the cold bulk metal. In other words, the
contact area remains cold as a result of a large mfp of phonons
compared with the contact side: the phonons have time enough to
depart and transfer their energy into the surrounding medium.

\begin{figure}[t]
\includegraphics[width=5.5cm,angle=0]{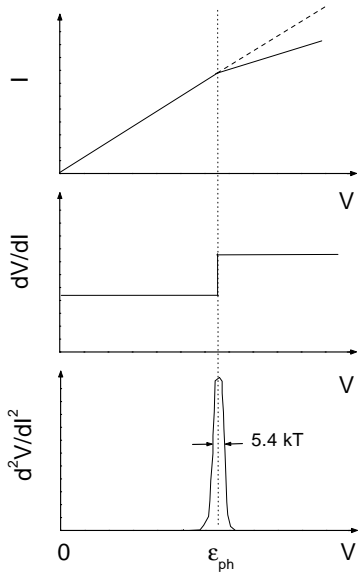}
\caption[]{\small Current-voltage characteristics $I(V)$, its
first and second  derivatives for PC of hypothetic metal in which
all phonons have the same energy $\varepsilon_{ph}$. The peak in
the second derivative will have the width 5.4$k_BT$ due to the
Fermi level smearing (see Chapter 8). } \label{fig6}
\end{figure}
\normalsize

For better understanding of scattering processes in PC, consider
the case of a certain hypothetic metal in which only the phonons
with the same energy $\varepsilon_{ph}$ are present. If the
electrons passing through the orifice have the excess energy $eV <
\varepsilon_{ph} $ they are unable to excite the phonons and the
electron trajectories are straight lines without scattering. The
contact resistance is precisely equal to the Sharvin's one, and
the current-voltage characteristics will be rectilinear
(Fig.\,\ref{fig6}). When the electrons obtain the energy
$\varepsilon_{ph}$, they begin to excite phonons, scatter, and
will not get into the orifice, so the current through the contact
will build up with voltage more slowly than in the absence of
scattering. It is clear that the current-voltage will have a break
at this point. The differential resistance $R_d =dV/dI$ dependence
will make a step, and the second derivative $d^2V/dI^2$ will
display a sharp peak at a specific energy (Fig.\,\ref{fig6}). If
the metal possesses several groups of phonons with definite
energies, then the $d^2V/dI^2$ curve will have a maximum for each
of them. In the general case at continued phonon spectrum the
second derivative of the current-voltage characteristics will be
similar to this curve.

As has been already mentioned, $d^2V/dI^2$ does not merely reflect
the phonon spectrum but is proportional to the probability
$G(\varepsilon)$ of emission of phonon with the energy
$\varepsilon$. The $G(\varepsilon)$ function is generally called
the PC spectral function of the electron-phonon interaction. In
the case of conductors, the transport and thermodynamic functions
of the electron-phonon interaction are also distinguished. The
former is connected with the current flow in metals, the latter
determinates the effective mass, velocity, and electron density of
states changes due to the electron-phonon interaction.

The PC spectral function $G(\varepsilon)$ introduced by analogy
with the above ones is a variety of transport function of the
electron-phonon interaction and differs by a stronger dependence
on the angle of electrons scattering in the contact. In the
orifice model this dependence is determined by the expression
$(1-\theta/\tan\,\theta)$, where $\theta$ is the scattering angle
of electron momentum. It is seen that at $\theta = \pi $, when the
electron reverses its motion direction this expression diverges,
i.e. such processes give maximum contribution into the PC spectrum
and are most significant. This essentially distinguishes the
situation in contacts from scattering in a homogenous conductor
resulting in normal resistance or from electron-phonon interaction
responsible for superconductivity. In the former case the
scattering efficiency increases with the angle as $(1-\cos
\theta)$, in the latter case it is independent of the angle at
all. Thus, all the functions in question can be written in the
form of Eq.(\ref{aF}), but with different $\alpha^2(\varepsilon)$
dependencies.

\section{Procedures for PCs creating}

We have already mentioned one of the techniques used to obtain
micro contacts in the tunneling structures. The schematic
representation of a contact with a short in the dielectric layer
is shown in Fig.7a. Such a conducting bridge often appears
spontaneously as a result of metallic dendrites intergrowth
through weak spots or defects of the dielectric film. The bridge
can be created artificially by means of mechanical loads causing
the oxide film damage or by means of electric breakdown. In the
former case use is made of a sharp needle creating high local
stresses, in the latter case the electric voltage is raised up to
the level exceeding the breakdown value at the weakest spot of the
dielectric. The latter case resembles electric microwelding when
metal is instantaneously melted in the constriction and is
purified from foreign inclusions. Certainly depending on the film
material, dielectric interlayer properties, magnitude and duration
of the current pulse, the degree of structure perfection and the
purity of metal in the constriction can vary in a wide range.
Using the film technology, one can obtain rather small PCs only of
a few nanometers with sufficiently high mechanical stability and
vibration resistance.

\begin{figure}[t]
\includegraphics[width=8cm,angle=0]{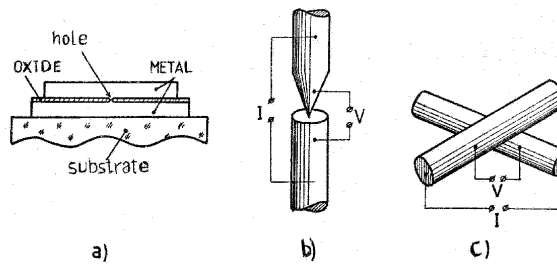}
\caption[]{\small Method of PC preparation   a) A film of metal is
deposited in vacuum onto a dielectric  substrate, a thin
dielectric layer is formed on the film by  oxidation or other
methods, on top of this layer the second  film of metal is
deposited. The technology is selected in  such a way that contacts
should be formed spontaneously in the process of preparation or as
a result of subsequent mechanical or electrical effects.
 b) "Needle-anvil" method: a specially processed metallic
electrode in the form of thin needle is slightly pressed to the
flat surface of the other electrode by means of  precision
mechanism. Here one can control the resistance or size of contact,
decrease or increase them.  c) Shear method - the contact is
formed by slightly  bringing together the edges of two electrodes
which can have random shapes and then by shifting one electrode
along  the other. As a result, contacts can be obtained in many
places and due to the shear stresses the metal deformations and
contaminations near the constriction are less than in the
preceding case.} \label{fig7}
\end{figure}
\normalsize

However, the fabrication of film PCs involves rather awkward
technique of vacuum deposition of metals. Besides, the structure
of metal in the films is usually less perfect than in the bulk
samples. That is why the creation of pressure contacts between
bulk electrodes gave an impetus to the development of research in
this field. A pressure contact of the needle-anvil type is shown
in Fig.7b. It is obtained by gradually approaching two electrodes
directly in liquid helium by means of precision mechanism.
Generally, the radius of curvature of the needle tip is a few
nanometers. Before measurement the electrodes are processed
mechanically, then chemically or electrochemically, to obtain
clean unperturbed surfaces of required shape. At this time the
surface is coated by a thin film of oxide or other non-conductive
compound. Due to the dielectric layer, at the first light touch of
electrodes the contact conductivity is often of the tunneling
character, i.e. the contact resistance decreases with the increase
of voltage. A metallic bridge appears at the increase of pressing
force either as a result of dielectric cracking or of breakdown by
the current pulse. The presence of dielectric film on the
electrode is necessary to stiffen the whole structure. In this
case the mechanical contact area is much greater than the size of
metallic constriction.

In the needle-anvil configuration, the metal in the contact area
is obviously deformed and polluted with pressed-in oxide residues,
as well as with various impurities absorbed on the surface. Better
results are obtained if the pressure contacts are created by the
technique shown in Fig.7c. In this instance the electrodes are
brought together till touching, then they are shifted relative to
one another in the plane of their crossing. As a result of it, the
deformation of metal here is much less while the oxide and surface
impurities are carried out of the contact area. Thus,
constrictions are formed in those places where the cleaned areas
of electrodes come in contact. With this technique, the samples
can take the form of small bulk cylinders or blocks of metal, as
well as films deposited on dielectric rods. By shifting the
electrodes independently in the mutually perpendicular directions
or rotating them about their axes, one can change the places of
contact and obtain a number of shorts with different resistances
within the same measurement cycle.

\begin{figure}[t]
\includegraphics[width=7.5cm,angle=0]{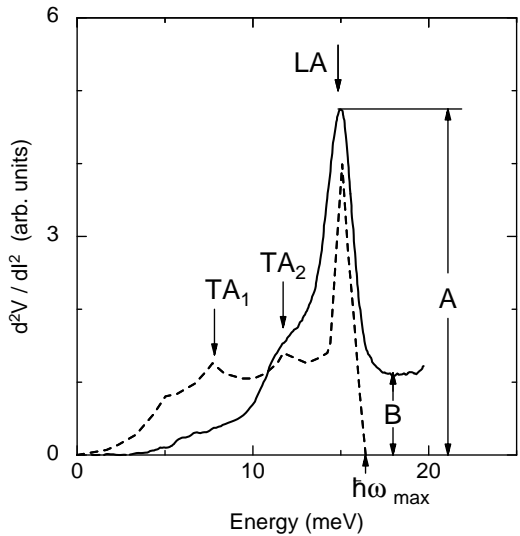}
\caption[]{\small Point-contact spectrum of one of simple metals,
Na (solid curve). The curve clearly reveals three features due to
the electron excitations of phonons in two transverse (TA) and one
longitudinal (LA) branches. Comparing the spectrum with the phonon
density of the states (dashed curve) one can see that in this case
electrons interaction with transversely polarized phonons is
considerably suppressed which is due to the closed Fermi surface
of this metal. } \label{fig8}
\end{figure}

The point-contact spectroscopy would be impossible if the
characteristics to be measured were critically dependent on the
constriction geometry which cannot be controlled due to the small
size of short. However, experience has shown that if the contacts
were chosen properly by means of experimental criteria, the
the reproducibility of PC spectra is not worse than in tunneling
spectroscopy where these choice is generally accepted. Let us
formulate the quality criteria of PCs which meet the requirements
of the theoretical clear orifice model, the orifice being much
smaller than the electron mfp. We shall take a PC spectrum of one
of simple metals, sodium (Fig.\,\ref{fig8}) as an example.

\normalsize

\begin{enumerate}
\item For pure metals, the contact resistance is within 1 to 100
$\Omega$ and the constriction has the metallic conductivity, i.e.
its resistance increases with the increase of voltage so that
$d^2V/dI^2 > 0 $ over the whole energy range.

\item Sharp peaks or other distinct features are observed in the
second derivative (PC spectrum) at the energies from zero to the
Debye one $\varepsilon_D \simeq\hbar \omega_{max}$ reproduced for
different contacts of the particular metal. For the energies $eV
>\varepsilon_D$ the spectrum assumes a certain constant value
called the background. The background parameter $\gamma =B/A$
(Fig.\,\ref{fig8}) should be small enough, different for each
metal, but as a rule not greater than 0.5.

\item Relative variation of differential resistance within the
spectrum should be maximum (from ones to tens per cent) at minimum
$\gamma$. The initial portion of the $d^2V/dI^2$ curve at the
energies $eV \ll \varepsilon_D$ is smooth and has no features.
\end{enumerate}

The above attributes make it possible, judging by the form of the
PC spectrum, to estimate the contact suitability for further
investigations of the electron-phonon interaction function.
Applying these criteria to the spectra shown in Fig.\,\ref{fig4},
one can see that they are satisfied by curves 1 and 2.

\section{How to measure the second derivative }

One of the widely used methods of measuring the current-voltage
characteristics derivatives is the modulation technique. It
consists in the following. Besides the measuring direct current
$I$, the sample is supplied with a definite frequency $\omega$ ac
current. In this case the contact voltage can be presented as a
Taylor series

\begin{eqnarray}
V(I+i \cos(\omega t))=~  V(I) +\nonumber \\ + \frac{dV}{dI} i
\cos(\omega t) + \frac{1}{2}\, \frac{d^2V}{dI^2}i^2 \cos^2(\omega
t) + \cdots  = V(I) +\nonumber  \\+\frac{dV}{dI} i \cos(\omega t)
+ \frac{1}{4}\, \frac{d^2V}{dI^2} i^2 (1+ \cos(2\omega t)) +
\cdots \label{vw} \label{Vw}
\end{eqnarray}

\begin{figure}[t]
\includegraphics[width=8.5cm,angle=0]{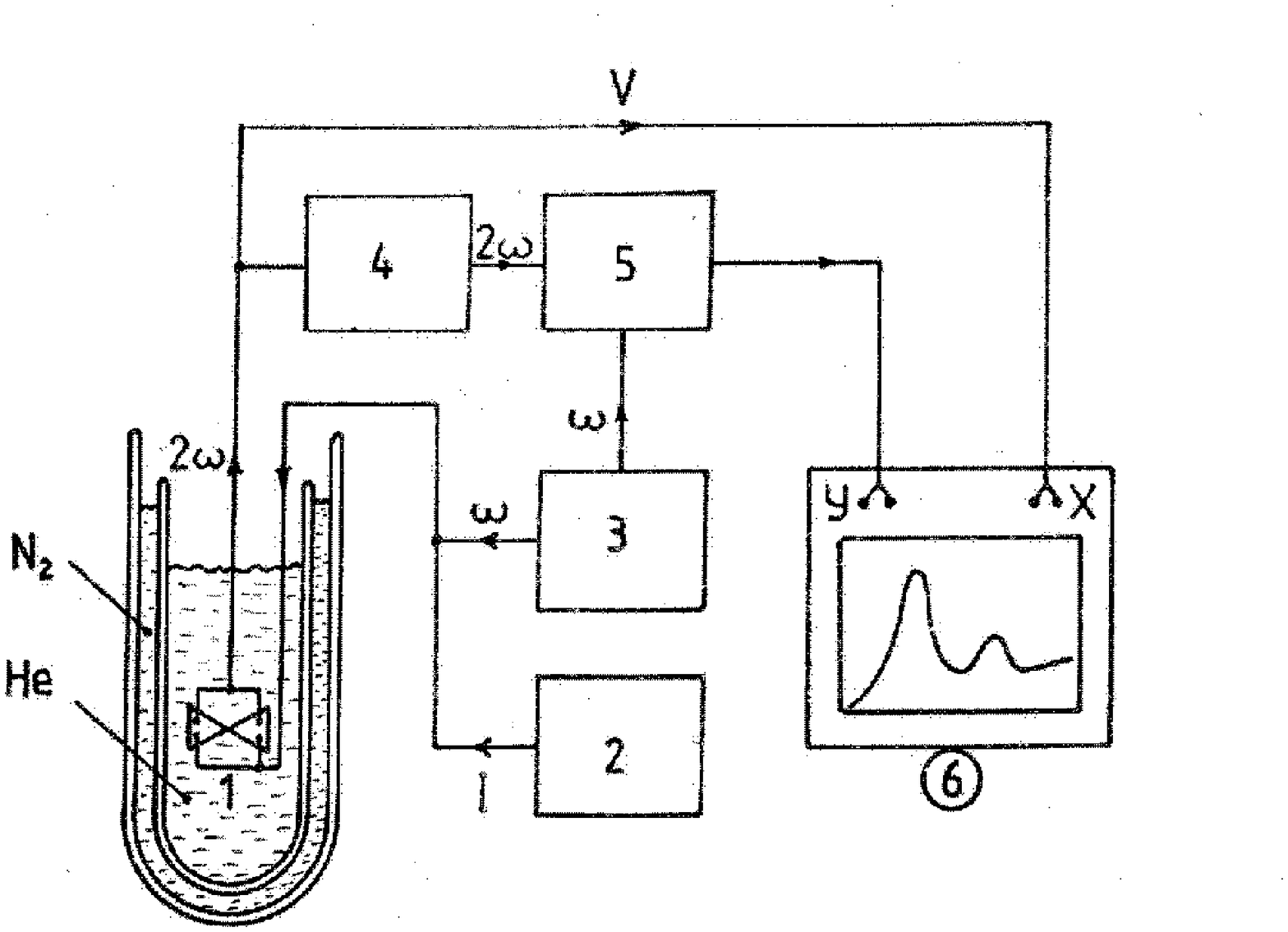}
\caption[]{\small Schema of the spectrometer: 1 - sample, 2 - DC
source, 3 - AC source, 4 - filter, 5 - look-in amplifier, 6 -
plotter.} \label{fig9}
\end{figure}

If the current $i$ is sufficiently small, the higher-order in $i$
terms can be neglected. It can be seen that the signal measured at
the frequency $\omega$ will be proportional to the first
derivative $dV/dI$ of the current-voltage characteristics, while
at the frequency $2\omega$ it will be proportional to the second
derivative $d^2V/dI^2$. The next problem is to measure this weak
voltage at the frequency $\omega$ and $2\omega$. This is achieved
by a spectrometer whose schema is shown in Fig.\,\ref{fig9}. The
sample (1) is held in a cryostat with liquid helium which is
immersed into an outer  cryostat with liquid nitrogen to reduce
the heat input and coolant evaporation. Connected to the sample
are four wires, a pair of which serves as an electric circuit,
through which the contact is supplied with direct current from the
source (2) and the alternating signal from the harmonic voltage
generator (3). The direct voltage $V$ created at the sample by
current $I$ is applied to the X-coordinate of the recorder (6) by
means of the other pair of wires. This pair is also used for
measuring the ac voltage of the second harmonic of modulating
signal which requires a special circuit for its amplification and
separation from noise. The circuit comprises the filter (4) and
lock-in amplifier (5). When registering this signal, proportional
to $d^2V/dI^2$, the filter is tuned to resonance to a required
frequency $2\omega$. The filter also serves for matching the
low-resistance sample with the high-resistance input of the
amplifier. Besides, the amplifier receives a $2\omega$ generator
signal used as a reference voltage. As a result of it, not only
the signal is amplified but also the required frequency is
separated. The voltage is converted into the direct one and
applied to the Y-coordinate of the recorder. Thus, when a linearly
increasing current from source (2) passes through the sample, the
recorder will automatically record the second derivative of the
current-voltage characteristics as a function of voltage.

\normalsize

As is seen from Eq.(\ref{Vw}) the signal at the frequency
$2\omega$ is proportional to the square of the modulating current
or first harmonic voltage amplitude $V_\omega$. Therefore, it
seems there are no special problems in measuring this signal by
merely increasing the alternating component $V_\omega$. However,
it cannot be made arbitrarily large as the ac voltage amplitude
determines the instrumental resolution of this technique. Exact
calculations indicate that an infinitely narrow spectral feature
when measuring $d^2V/dI^2$ by the modulation technique will be
smeared to form a bell-shaped maximum $1.7V_\omega$ wide. In real
experiments, the value of $V_\omega$ is as a rule not greater than
a few millivolts, and in separate cases it can be equal to several
tens of millivolt. In this case the value of measured signal which
characterizes the nonlinearity of current-voltage characteristics
associated with the electron-phonon interaction is of the order of
microvolts. At the same time the measuring circuit sensitivity is
two orders of magnitude higher which makes it possible to reliably
register the PC spectra.

It should be also noted that the point-contact spectroscopy
resolution is dependent on the temperature $T$ due to the Fermi
level smearing by about $k_BT$. Therefore, to obtain distinct
spectral features, the condition $k_B T \ll \varepsilon_D$ should
be satisfied, i.e. measurements should be taken at temperatures
considerably lower that the Debye ones. Under the action of the
instrumental and  temperature factors, the resulting resolution is
expressed by
\begin{equation}
\delta = [(5.4 k_BT/e)^2 + (1.7 V_\omega)^2]^{1/2} \label{smear}
\end{equation}

Generally, measurements are taken in liquid helium at a
temperature of 1.5 to 4.2 K and modulating signal of 0.3 to 1.5
mV. As a result of it, the resolution calculated by
Eq.(\ref{smear}) is 0.9 to 2.9 mV or, expressed in degrees, 10 to
34 K. This is at least an order of magnitude below the Debye
temperatures of typical metals which enables obtaining a
sufficiently detailed spectrum of electron-phonon interaction.

\section{Point contact instead of reactor}

Obviously, the headline of this section is rather advertising than
true to life. A long time will pass till a nuclear reactor can be
replaced by a competitive source of energy or research tool. Our
headline is used in the sense that PC can compete with the reactor
in certain fields of solid-state physics. We shall consider the
experimental techniques of phonons investigation in metals and the
determination of phonon spectrum.

One of the principal methods used for determining the phonon
dispersion law $\varepsilon (q)$ is the study of neutron
scattering in a crystal. The source of neutrons is usually a
reactor. For this purpose use is made of the so-called thermal
neutrons with the energy of 30 meV or 3000 K and, respectively,
the de Broglie wavelength comparable to the lattice constant.
Knowing the momentum and energy of incident neutrons, then
measuring their energy and momentum after scattering, on the basis
of conservation laws one can determine the energy and momentum of
phonon by which the neutron was scattered. Thus one obtains the
dispersion curves shown in Fig.2a. Since the law of phonons
dispersion in a crystal is anisotropic, the dispersion is more
often measured only for few principal crystallographic directions
as the measuring procedure itself is rather long and requires a
powerful source of neutrons. To obtain a phonon spectrum, one
needs the dispersions in the lattice, then one should calculate
all the phonons whose energy falls within the interval from
$\varepsilon_k$ to $\varepsilon_k+\Delta \varepsilon =
\varepsilon_{k+1}$ (where $\varepsilon \leq \varepsilon_D$ and
$\Delta \varepsilon \ll \varepsilon_D$), then plot this value
(Fig.2b). Then repeat this procedure for the next interval from
$\varepsilon_{k+1}$ to $\varepsilon_{k+1}+\Delta\varepsilon$, etc.
The less the interval $\Delta \varepsilon$, the more accurate is
the $F(\varepsilon)$ curve. But, as mentioned above, it takes a
long time to measure the dispersion curves, and they are available
only for several principal crystallographic directions in the
lattice. To obtain the dispersion law for a maximum number of
directions, use is made of theoretical calculations describing the
lattice thermal vibrations. The formulas employed for this purpose
include the parameters which have definite values for one or
another solid. These parameters are found by fitting the
theoretically calculated dispersion curves to the experimental
ones. The obtained parameters make it possible to calculate the
dispersion curves for any direction and, therefore, to calculate
the energy of phonons with any momentum value both in magnitude
and direction, and to plot the $F(\varepsilon)$ curve. The
required calculations are rather bulky, voluminous and need the
use of a powerful computer. Besides, the result depends on the
model selected, i.e. on how accurately the theory takes into
account the real processes of atoms interaction in the lattice. In
principle, it is possible to obtain the information on the phonon
density of states in a more direct "modelless" way. For some
metals one can use the one-phonon incoherent scattering of
neutrons when the directly measured scattering cross-section of
neutrons with a specific energy is proportional to the phonon
density of states. However, the latter technique is not suitable
for all metals and does not have high resolution.

How can the point-contact spectroscopy be of any help here and how
can it substitute the reactor? We have demonstrated that the form
of electron-phonon interaction function $G(\varepsilon)$ is
dictated by the phonon density of states. It turn out that for
practically all metals under investigation the spectrum extrema
are stipulated by their presence in $F(\varepsilon)$. Moreover,
for some metals, mainly polyvalent ones, such as Pb, In, Sn, Mg,
Be, etc., these two dependencies $G(\varepsilon)$ and
$F(\varepsilon)$ practically coincide in form. Therefore, the
point-contact spectroscopy can be referred to as a new method of
phonon spectroscopy in metals. It should be mentioned that the PC
experiments are simple and accessible for any physical laboratory.
One should only have liquid helium for low temperature
measurements and ordinary electronic engineering equipment. It is
noteworthy that PCs can be created by means of very small amounts
of a substance, a few cubic millimeters of which being sufficient.
It becomes especially important for studying rare substances
available only in small quantities.

\begin{figure}[t]
\includegraphics[width=8cm,angle=0]{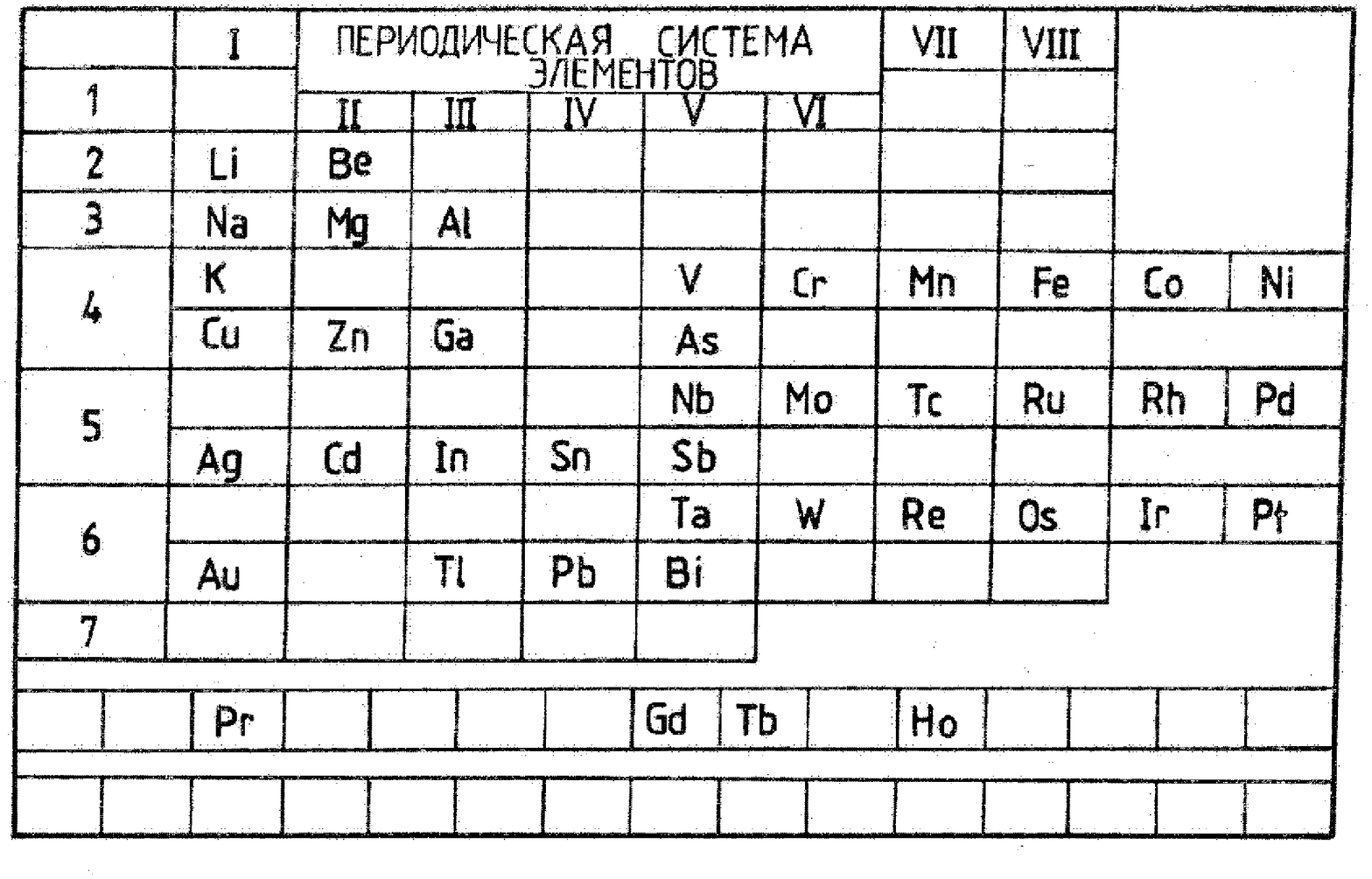}
\caption[]{\small Mendeleev's periodic table containing only those
elements which have been studied by point-contact spectroscopy
(until 1988).} \label{fig10}
\end{figure}
\normalsize

To date, the PC spectra have been obtained for almost four tens of
pure metals indicated in Fig.\,\ref{fig10}. If we drop the
rare-earth elements, we have to measure yet of about ten metals,
basically highly active alkali-earth and alkali ones. By means of
point-contact spectroscopy, it became possible to refine the
positions of extrema and other features of the $F(\varepsilon)$
function. For a number of metals it was possible to determine the
shape of the whole curve, i.e. to find out the number of maxima,
their positions with respect to energy and relative intensities.

\section{Study of electron-phonon interaction}
It has already been mentioned that the spectral function
$G(\varepsilon)$ is a kind of transport function of the
electron-phonon interaction. By comparing $G(\varepsilon)$ with
phonon spectra for various metals, one can study the Fermi surface
effect on the dynamics of electron-phonon interaction. Thus, in
alkali metals Na and K the Fermi surface is practically a sphere.
At such an isoenergetic surface, it is difficult for electrons to
interact with transversely polarized phonons. In fact, the
features associated with scattering by transverse (T) phonons are
strongly suppressed in the PC spectra of sodium (Fig.\,\ref{fig8})
and potassium. On the contrary, the Fermi surface in noble metals
is open which is favorable for interaction with transverse
phonons, so in the PC spectra of copper (Fig.\,\ref{fig11}), gold,
silver the peaks corresponding to these phonons prevail. Finally,
in the polyvalent non-transition metals: lead, tin, magnesium,
berillium (Fig.\,\ref{fig11}), etc., the Fermi surface is of
intricate configuration and few phonons of one or another
polarization are released. In this case the $G(\varepsilon)$
function shape is closed to the dependence of phonon densities of
state, i.e. one can assume that the parameter $\alpha^2
(\varepsilon)$ in the formula (3) is weakly dependent on energy.

On the other hand, the electron-phonon interaction function itself
is of fundamental importance. It determines the properties of
metals stipulated by the electron-phonon interaction, in
particular the possibility of superconducting transition, as well
as various kinetic parameters characterizing the charge and energy
transfer. The point-contact spectroscopy yields detailed
information on the relative force of electrons binding with some
groups of phonons, e.g. with those transversely or longitudinally
polarized. Thus, it becomes clear which phonons play more
important role in various processes. For normal
non-superconducting metals, the study of PC spectra is the only
experimental technique which enables direct measurement of energy
dependence of the spectral function of electron-phonon
interaction.

\section{Magnons and excitons}

From the very childhood we have used to metals breaking down into
magnetic and nonmagnetic ones without thinking on the nature of
such difference. This classification of metals has been explained
in terms of quantum theory. Generally speaking, magnetism is a
purely quantum-mechanical phenomenon that cannot be explained in
the framework of classical theory. As one of the well-known
physicist said, if the Planck constant had gone to zero, there
would have been no science of magnetism.

As is known from the atomic physics, an electron has its moment of
momentum, the spin. Besides, an electron in an atom has an orbital
magnetic moment due to the motion about the nucleus. To determine
the total magnetic moment of an atom, one should consider its
electronic structure. If the atoms with a magnetic moment are
united in a lattice, it is not enough to make the whole material
magnetic. Everything will depend on the actual interaction of
moments in which the so-called exchange forces play an important
role. These forces arise due to quantum indistinguishability of
particles and the Pauli principle. The exchange forces are not
connected with direct interaction of spin magnetic moments which
is weak but are dependent on their relative positions. The
exchange interaction just makes the magnetic moments of separate
atoms feel one another and line up in a certain way: in parallel
or in antiparallel direction. In the former case we obtain a
ferromagnet, in the latter case antiferromagnet.

Consider, for example, a ferromagnetic metal in which the magnetic
moments of atoms have the same direction. Let for some reason, for
example due to thermal vibrations, one atom reverse the direction
of its moment. Then the neighboring atom will be affected by a
force tending to change over the direction of its magnetic moment
too, etc. Hence, a wave of magnetic moment reversals will pass
through the crystal. This wave was called the spin wave. The spin
reversal events in magnetic materials are usually considered in
the language of spin waves convenient for the theory. According to
the quantum mechanical principle of dualism, the waves can be
substituted by quasi-particles and, by analogy with a phonon, the
spin oscillation wave was called a magnon.

Thermal vibrations "rock" the magnetic moments and with the
increase of temperature the number of atoms with reversed spins,
or, in other words, the number of magnons increases. When the
energy of thermal vibrations becomes equal to that of exchange
interactions, there will be no ordering of magnetic moments and
the magnetic metal will become nonmagnetic. The temperature at
wich it occurs was called the Curie temperature $\Theta_{c}$. The
latter is similar to the Debye temperature for phonons: there can
be no magnons with the energy above $k_B\Theta_c$ in metal.

Magnons are quasi-particles similar to phonons. They carry heat,
scatter the conduction electrons, i.e. affect the kinetic
properties of metals. Therefore, it is quite reasonable to suppose
that the electrons scattering by magnons or, more precisely, the
magnons excitation by electrons in a PC should cause non-linearity
of current-voltage characteristics, as is the case with phonons,
and the PC spectrum will display the features associated with the
electron-magnon interaction. Indeed, such features were registered
when investigating the PC spectra of rare-earth metals Gd, Tb, Ho
which are ferromagnetics at low temperatures. A question arises
why rare-earth metals were investigated but not less exotic
well-known ferromagnetics, such as Fe, Ni, Co. The matter is that
in these metals the Curie temperature, and consequently the energy
of magnons, is considerably greater than those of phonons. At this
time it turns out that the electron-phonon interaction essentially
decreases the electrons mfp and the ballistic regime in a contact
is disturbed before reaching the energy of characteristic magnon
frequencies. On the contrary, in the above mentioned rare-earth
metals the Curie temperature is less than the Debye temperature
and the electron-magnon interaction features show up at lower
energies than the phonon ones.

In a solid, the conduction electrons can also interact with other
elementary excitation or "-ons". Let us dwell on one of such
quasi-particles, an exciton, representing the electronic
excitation in a crystal of semiconductor or dielectric. As a
result of interatomic interaction, an exciton propagates over a
crystal in the form of a wave without carrying charge or mass.
Excitons are most intensively emitted from semiconductors and
dielectrics where they present a bound state of the conduction
electron and hole. However, there exist some exciton-like
excitations in metals as well.

\begin{figure}[]
\includegraphics[width=6.5cm,angle=0]{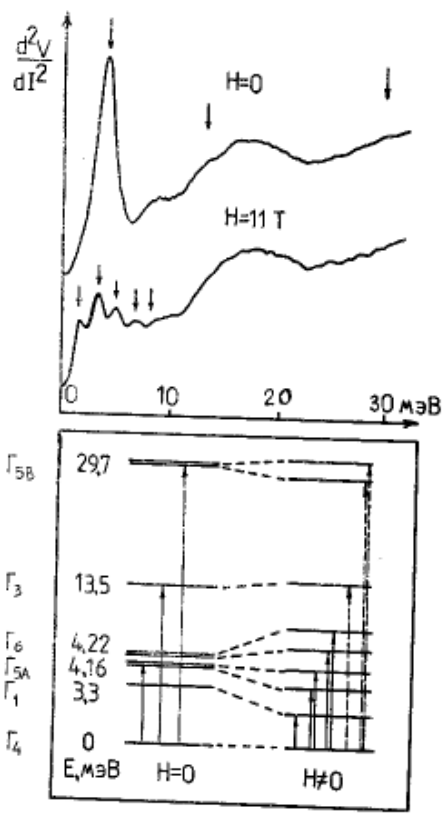}
\caption[]{\small Above: Point-contact spectrum of PrNi$_{5}$ in a
zero magnetic field and in a field of 11 Tesla. Arrows indicate
the peculiarities due to the Pr$^{3+}$ ion transitions between the
energy levels split by the intracrystalline field.  Below: scheme
of the Pr$^{3+}$ energy levels in the intracrystalline electric
field and their additional splitting in the external magnetic
field. Arrows indicate the allowed transitions marked on the PC
spectra. The transitions marked by dashed lines are weakly
probable in the magnetic field and are not visible on the spectra.
} \label{fig12}
\end{figure}

Thus, in the compound PrNi$_5$ the main nine-fold degenerated
state of Pr$^{3+}$ ion is split by the intracrystalline electric
field of lattice atoms into six energy levels, three of which
remain doubly degenerate, the other three presenting a separate
level of energy (see diagram in Fig.\,\ref{fig12}. When the
Pr$^{3+}$ ion is excited, it can reside in one of the higher-lying
states, for example $\Gamma_{5A}$, in which it has a magnetic
moment. Such an excited state is transmitted from ion to ion, i.e.
it can be treated as a magnetic exciton. The processes of
Pr$^{3+}$ ions excitation by electrons should obviously cause the
PC spectrum features at corresponding energies.

\normalsize

At low temperatures, the Pr$^{3+}$ ion is in the ground state
$\Gamma_{4}$ with minimum free energy. When the ion is excited, it
can move to the above-mentioned level $\Gamma_{5A}$, as well as to
the higher-energy levels $\Gamma_{3}$ and $\Gamma_{5B}$. Other
transition to the levels $\Gamma_{1}$ and $\Gamma_{6}$ are
forbidden by the selection rules. However, in an external magnetic
field this exclusion is lifted (it remains only for one
high-symmetrical direction along the hexagonal axis), and the
Zeeman splitting of spin-degenerate levels occurs (see
Fig.\,\ref{fig12}).

The processes of Pr$^{3+}$ ion excitation by electrons, as a
result of which the ion moves to a higher-lying level, bring about
their inelastic scattering and manifest themselves in the PC
spectrum as peaks at the corresponding energies. In the zero field
the PrNi$_{5}$ spectrum distinctly reveals three features
associated with the transitions shown in Fig.\,\ref{fig12}(lower
panel). The most intensive peak corresponds to the Pr$^{3+}$ ion
excitation to the state $\Gamma_{5A}$. Evidently, these processes
will make maximum contribution into the resistance of this
substance. The probability of other transitions and, respectively,
their influence on electric conductivity and other kinetic and
thermodynamic properties of PrNi$_{5}$ is appreciable less. These
are the simplest qualitative conclusions that can be made directly
from the form of PrNi$_{5}$ PC spectrum.

The probability of transition to previously forbidden levels
manifests itself in the magnetic field and Zeeman splitting of the
remaining degenerate levels occurs. Consequently, the PC spectra
should change when the magnetic field is switched on,which can be
seen in Fig.\,\ref{fig12}(upper panel). An especially noticeable
evolution occurs with most intensive maximum. Here, together with
the peak splitting at 4 meV, one can see additional features
connected with the transitions to the levels $\Gamma_{6}$ and
$\Gamma_{1}$. Hence, instead of one maximum several maxima appear
in the magnetic field. Note that these features can be distinctly
observed only in a sufficiently high magnetic field. Thus, for
example, the amount of splitting evaluated by the formula $E=2g\mu
H$ (where $g$ is Lande factor, $\mu$ is the Bohr magneton) should
be greater than the resolution of this technique which is about
1~meV. As a result of it, the magnetic field amplitude should
exceed 4-5~Tesla and the higher the amplitude, the more distinct
are the features. That is why PrNi$_{5}$ was studied in a field of
up to 21~Tesla, while the figure, by way of example, shows the
spectrum for an intermediate field.

It should be noted that measurements were carried out on
single-crystalline samples of PrNi$_{5}$ for three principal
crystallographic directions, which made it possible to detect also
the anisotropy of Zeeman splitting. When doing so, we for the
first time obtained the experimental data on the exact positions
of levels the transitions to which are forbidden in zero field.
These data were used for comparison with theoretical calculations
which allowed us to determine the parameters characterizing the
intracrystalline field in a particular compound. The above results
can serve as an example of how new data were obtained by
point-contact spectroscopy which enabled the construction of a
quantitative model of the intracrystalline effects influence on
the PrNi$_{5}$ properties.

\section{From pure metals to alloys and compounds}

In principle, any energy-dependent mechanism of electrons
scattering in metals can be studied by means of point-contacts. To
put it in another way, the processes influencing the inelastic mfp
of electrons in the lattice will be reflected in the point-contact
spectra. This can be easily proved using the already mentioned
Wexler formula (2) for the contact resistance. The second term in
this formula, the Sharvin resistance, is a constant for the
product of $\rho l$ is constant. The Maxwell resistance can be
rewritten as $R_M = \rho l/d~(1/l)$ where the first cofactor is
also constant and the second $1/l$, is energy- dependent if there
exist the electron scattering processes affecting the electron
mfp, i.e. $l = l(\varepsilon) = l(eV)$. For the second derivative
of the current-voltage characteristics we shall write $d^2V/dI^2
\sim R~dR/dV$. Substituting instead of $R$ its expression from
Eq.(\ref{Rwex}) we obtain that
\begin{equation}
      d^2V/dI^2 \sim \rho l/d \,{\rm d/d}V [1/l(eV)] \label{d2v}
\end{equation}
Thus, the second derivative will have extrema at those energies at
which the mfp of electrons changes rapidly. In the case of
electron-phonon scattering the electron mfp considerably decreases
at those energies at which there exist a large number of phonons,
i.e. at the phonon spectrum maxima. That is why a peak can be
expected in this place of $d^2V/dI^2$.

It has been previously noted that the spectral regime is realized
only in very small contacts, such that the electron mfp should
exceed the constriction size. There are two mfp of electrons in
conductors: the elastic mfp $l_e$ after passing which the electron
changes its momentum or direction of motion, and the inelastic mfp
$l_i$ at which the electron changes its energy, $l_e$ is
independent of the excess energy of electrons and, according to
Eq.(\ref{d2v}), should not affect the point-contact spectrum form.
A question arises whether the point-contact spectroscopy is
possible if the elastic mfp becomes less than the short size. This
question is not idle since many alloys and compounds have rather
small $l_e$. The positive answer makes it possible to apply the
point-contact spectroscopy technique not only to pure metals whose
number, though large, is limited, but also to practically
unlimited collection of more or less complex alloys or compounds
of different metals. It turned out that the technique works under
less strict conditions, namely if the inelastic relaxation length
during the diffusion motion of electrons in the contact $l_D$
which in the case of alloys is determined by $l_D=(l_e
l_i/3)^{1/2}$ is greater than the size of contact. The condition
$d \ll l_{D}$ can in many cases be satisfied for PCs made of
complex substances and the spectral information on electron
scattering mechanisms can be obtained. The regime at which the
condition $l_{e} \ll d \ll l_{D}$ is met was called the diffusion
regime. In this case, as was shown by theorists, the
proportionality between $d^2V/dI^2$ and the spectral function
$G(\varepsilon)$ is preserved, only in the right-hand part of
Eq.(\ref{aF}) an additional factor appears that is proportional to
the $l_e/d$ ratio and leads to the decrease of point- contact
spectrum intensity, for $l_e/d \ll 1$.

If impurities are introduced into a perfect lattice, it will cause
the decrease of elastic electron mfp $l_i < l_e, d$, which is the
smaller, the higher concentration of the impurity. At this time
the point-contact spectra will not change at low impurity
concentrations (up to several per cent), only their absolute
intensities drop. A more interesting case is when the mass $m_i$
of the impurity atom substantially differs from the mass $M$ of
lattice atoms. Thus, for example, if $m_i \ll M$, there arise the
so-called local oscillations of the impurity atom with a frequency
higher than the maximum one of the host lattice vibrations. At the
inverse relation $m_i \gg M$, the quasi-local oscillations with a
frequency essentially lower than the Debye one appear in the
spectrum of lattice frequencies. It can be readily understood
qualitatively if we recall that the frequency of oscillator
(pendulum) oscillations is inversely proportional to the square of
its mass. It appears that the electrons scattering by local and
quasi-local oscillations occurs inelastically, i.e. with the
change of energy. Hence, the peculiarities connected with such
oscillations can be expected in the point-contact spectra.

\begin{figure}[t]
\includegraphics[width=7cm,angle=0]{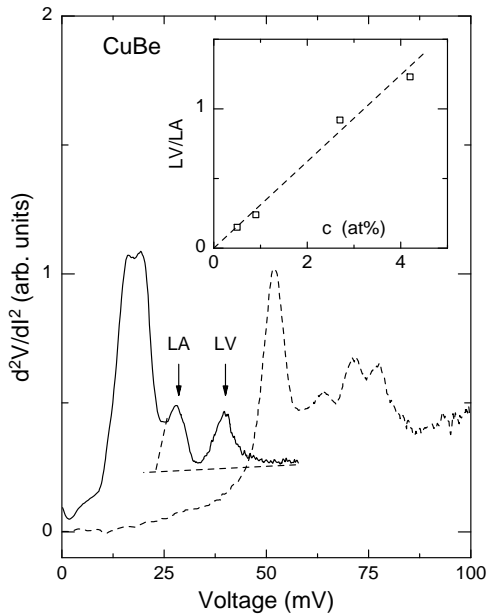}
\caption[]{\small Point-contact spectrum of copper alloy with 4.2
at.\% of beryllium. It is seen that in the Cu spectrum (between 0
and 30 mV) has developed a new maximum (LV) at 40 mV which is
absent both in the spectrum of pure copper and beryllium (dashed
curve). It is connected with local vibrations of the Be atoms in
the Cu lattice. Inset shows increasing of LV maximum intensity
relative to LA phonon peak in Cu with beryllium content rise.}
\label{fig11}
\end{figure}

Fig.\,\ref{fig11} shows the spectrum of beryllium, as well as the
CuBe alloy containing 4.2 atomic percent of Be. One can see that
an additional feature has appeared in the $d^2V/dI^2$ curve. This
feature cannot be found either in the spectrum of copper or in
that of beryllium. This is exactly the band of local oscillations.
In this case the impurity atoms are seven times lighter than the
copper ones. Another example of an alloy with quasi-local
oscillations is magnesium with the admixture of lead atoms. Here
the difference in masses is 8.5 times. The point-contact spectra
of this alloy also reveal a feature that could be expected in the
region of energies essentially lower than the Debye one.

\normalsize

The other impurities which strongly scatter the electrons are the
atoms of rare-earth metals of iron group which retain their
magnetic properties when added to a normal metal. The electrons
scattering by such elementary magnetics in a bulk conductor causes
an abnormal for metals rise of resistance at low temperatures
known as Kondo effect. PC has proved to be a very sensitive
indicator for detecting magnetic impurities in metal. Thus, when
adding only 0.01$\%$ of iron or manganese atoms to copper, the
point-contact spectrum will display distinct additional features
in the energy range 1-2 meV which are connected with the electrons
scattering by localized magnetic moments. One can evaluate the
number of impurity atoms getting into the contact area at the
above concentration $c=10^{-4}$. The total number of atoms in a
constriction with size $d$ will be $V/\theta$, where $V \simeq
d^3$ and $\theta$ is the volume per atom. The number of impurity
atoms $N_i$ is estimated as a value of the order of $cV/\theta$.
For copper $\theta = 1.17 \times 10^{-23} $cm$^3$, and the size of
PCs in question is about 1 to 2 $\times 10^{-6}$ cm. Therefore, we
obtain the value of $N_i$ equal to 8-70 atoms. It is of interest
that the PC area contains a meagre number of atoms whose effect
is, however, noticeable in the measured characteristics. Thus, we
can say that the point-contact spectroscopy is a unique method of
studying the electrons scattering by single microscopic centers.

\section{Contact heating}

\begin{figure}[t]
\includegraphics[width=7cm,angle=0]{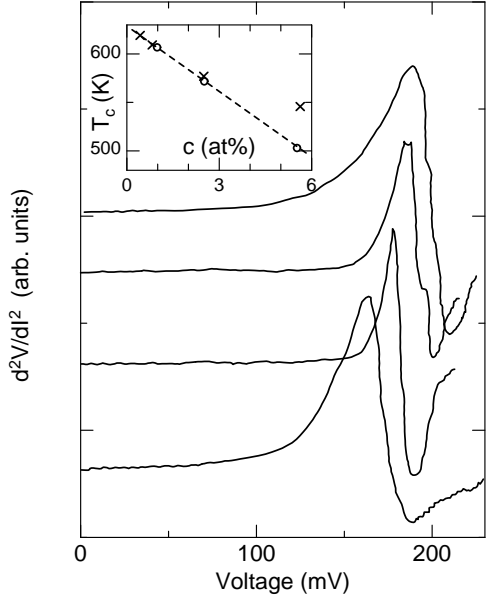}
\caption[]{\small Second derivative for PCs of nickel (upper
curve) and its alloys with beryllium  Ni$_{1-x}$Be$_{x}~x\simeq
0.01, 0.03, 0.06 $ (bottom curves) far above the phonon energies.
The observed sharp features are due to the near-contact metal
transition from the ferromagnetic state to the paramagnetic one
when heating by current up to the Curie temperature. The shift of
features towards smaller voltages (see inset) when adding
beryllium to nickel is due to the Curie temperature decrease in
the alloy and proceeds according to Eq.(\ref{T-V}).}\label{fig13}
\end{figure}

So far we consider PCs whose dimensions are small in comparison
with the electron mfp. As has been shown, in this case a highly
nonequilibrium function of electrons distribution is realized and
there exist two groups of electrons whose energies differ by $eV$.
Relaxation of such a nonequilibrium state leads to the excitation
of quasi-particles with the energy $\varepsilon \leq eV$ and to
the nonlinearity point-contact current-voltage characteristics
which contains spectral information about the energy distribution
of quasi-particles under investigation. It turned out that even in
the case of small elastic mfp the spectroscopic capabilities of
contacts remain. For this purpose it is only necessary that the
diffusion length of energy relaxations should be greater than the
size of constriction. It is of interest what will happen if the
elastic and inelastic lengths will be considerably smaller than
the size of contact. In other words, what will happen if the
contact dimensions are large or when studying the substances with
sufficiently short relaxation lengths. It appears that in such
constrictions the flow of current of the required density leads to
conventional Joule heating. The heat removal from such large
contacts with small mfp is less efficient than in the ballistic
regime. Since in metals the heat removal from the hot region to
the bulk of the sample is principally effected by the conduction
electrons, it can be shown that when the Wiedemann-Franz law
holds, the voltage drop at the contact and the metal temperature
at the center $T_{pc}$ are related as
\begin{equation}
     (k_BT_{pc})^2 = (k_BT_0)^2 + (eV)^2/4L \label{T-V}
\end{equation}
where $T_0$ is the measurement temperature, $L$ is the Lorentz
number. If measurements are taken at low temperatures, i.e. $T$=0,
the contact temperature will be linearly dependent on its voltage
$k_BT_{pc} = eV / 2\sqrt{L}$. Substitute the known Lorentz number
and obtain that $T_{pc} \simeq$ 3.2 K/mV, i.e. a voltage of 1 mV
raises the temperature at the contact center by 3.2 degrees. The
situation seems paradoxical: the samples are at a temperature of
liquid helium, but if we apply to the contact a voltage of only
100~mV or 0.1~V, the temperature of the contact becomes higher
than the room temperature. Is it possible and how to verify this
conclusion? Yes, it is, and rather simply. If we take the
above-mentioned ferromagnetic metals, at the Curie temperature the
transition from ferromagnetic into paramagnetic state occurs. As a
result of it, an inflection appears in the electrical resistivity
curve, i.e., magnons and associated scattering disappear. As noted
above, the contact resistance is expressed by the Maxwell formula
$R_M = \rho /d$, so when reaching the Curie temperature in the
contact, its resistance curve will have features similar to that
of $\rho (T)$. These peculiarities, which must show up still
definitely in the second derivative of current-voltage
characteristics $d^2V/dI^2\simeq dR/dV$. Such features of the
point-contact spectrum for ferromagnetic metals were observed at
voltages equal, for example for Ni, to about 200~mV. This
corresponds, according to Eq.(\ref{T-V}), to the known Curie
temperature of nickel equal to 630 K (Fig.\,\ref{fig13}). When
adding beryllium  atoms to Ni, $T_{c}$ of this alloy decreases and
the feature of $d^{2}V/dI^{2}$ also shifts to the region of lower
voltages in full agreement with the above formula. This testifies
of the fact that, actually, the temperature reached at the contact
center causes the metal transition from the magnetic to the
nonmagnetic state.

\normalsize

The criterion of thermal regime is also the dependence of the
above anomaly position on the bath temperature $T_0$. At this time
it can be observed without making measurements in helium. The
anomalies were found out for nickel contacts at room temperature,
only they were shifted towards smaller voltages. It follows from
Eq.\,(\ref{T-V}) that the required temperature in the contact is
reached at the lower voltages, the higher the measurement
temperature $T_0$. Thus, all kinds of $d^2V/dI^2$ anomalies, if
they are of thermal nature, will shift over the voltage scale
depending on the measurement temperature. At the same time the
spectral features of $d^2V/dI^2$ do not change their positions
with temperature, they only broaden according to
Eq.\,(\ref{smear}).

One of the straightforward and easy-to-grasp examples of the
experimental evidence of temperature increase inside the contact
was its critical voltage observation. This critical voltage cannot
be exceeded no matter how the current rises. The nature of this
phenomenon consists in that each metal has the so-called
"softening" temperature, about one-third of the melting one, at
which the metal readily lends itself to plastic deformation. When
the contact voltage reaches the corresponding "softening" value,
under the action of pressing force the contact spot will increase
or the contact resistance will decrease, so that at the current
build-up a certain critical voltage cannot be exceeded. This was
observed for many metals with different melting temperatures in
which they correlated with the contact critical voltages.

In the above thermal regime, the current-voltage characteristics
nonlinearity is governed by the temperature dependence of
resistivity, as the contact resistance is $\rho(T)/d$ and its
temperature is voltage-dependent. This enables obtaining the
$\rho(T)$ behavior by low-temperature measurements of the contact
resistance as a function of voltage, i.e. without direct
measurements of $\rho(T)$. The advantages of this technique are
the possibility of direct measurements $d\rho(T)/dT$ derivatives
proportional to $dR/dV$ or $d^2V/dI^2$, the possibility of using
small quantities of substances, the absence of need in the systems
for controlling and stabilizing the sample temperatures. The above
technique was called modulation temperature spectroscopy since
when measuring the current-voltage characteristics the sample is
supplied with a low modulation voltage and the contact temperature
in the thermal limit is changed (modulated) in step with the
voltage. Note that in this case the contact size has no upper
bound and the result is independent of the constriction geometry.
It is only necessary that the heat should be removed from the
contact mainly by electrons.

Another convincing proof of metal heating inside the contact in
the thermal regime was the detection and explanation of
current-voltage characteristics asymmetry in a number of
heterocontacts depending on the applied voltage polarity. Consider
an electric circuit comprising a heterocontact formed between
different metals and a potentiometer connected to it by two wires.
Let the sample temperature be $T_0$, the instrument is kept at
room temperature $T_{room}$ and the wires have the temperature
gradient from $T_{room}$ to $T_0$. This will cause the electron
diffusion from a hotter end to a colder one and the appearance of
thermoelectromotive force \footnote{In real metals the
thermoelectric phenomena are connected not only with the electron
diffusion but also with more complicated processes, such as phonon
drag, etc., which should not be necessarily taken into account in
this consideration.}. However, the gradient exists only in wires
which are identical and equal quantities of electrons per unit
time will arrive at the cold ends of wires. The potentials will be
the same and the outside device will read the voltage (potential
difference) equal to zero. The situation will change if the
contact region will heat up and the temperature will become other
than $T_0$. A similar diffusion of electrons from the hot region
to the cold one will start in metals making up the heterocontact.
As the metals are different, in one of them the diffusion will
proceed faster than in the other. As a result of it, the
potentials at cold ends will differ and the device will read some
potential difference due to thermoelectromotive force. To describe
the thermoelectric phenomena quantitatively, we will introduce the
Seebeck coefficient $S$. It indicates, for example, what potential
difference will exist at the ends of a conductor if a temperature
difference of one degree is maintained between them. If one knows
$S$ of each metal in a heterocontact, one can calculate the
potential difference between the cold ends of metals making up the
heterocontact. The potential at one end will be
$S_{1}(T_{PC}-T_{0})$, at the other end it will be
$S_{2}(T_{PC}-T_{0})$ and the potential difference registered by
the outside instrument will be $(S_{1}-S_{2})(T_{PC}-T_{0})$. To
be more precise, we shall note that the above considerations are
valid if the coefficient S is temperature- independent. Otherwise,
the sample under study should have been divided into small
sections at which the temperature difference is sufficiently small
to treat $S$ as a constant, and then to sum up the thermoemf of
all these sections. This actually implies integration and finally
one can write the well-known formula
$V_{temf}=\int_{T_{0}}^{T_{PC}}(S_{1}-S_{2})dT$ to determine the
thermoemf in a contact of two metals.

What will be the effect of thermoemf when measuring point-contact
characteristics of heterocontacts in the thermal regime? The
answer is rather obvious: it causes the asymmetry of
current-voltage characteristics depending on the polarity of
voltage applied to the contact. The reason of it is as follows. In
one polarity denoted by, say (+), the voltage applied to the
contact will coincide with $V_{temf}$ and add to it while in the
other polarity denoted by (-) the applied voltage will be
subtracted. As a consequence, at current $I$ through the contact,
its voltage in the former polarity will be
$V^{(+)}=IR_{PC}+V_{temf}$, and in the latter polarity
$V^{(-)}=IR_{PC}-V_{temf}$, where $R_{PC}$ is the contact
resistance at zero voltage. Thus, the resistance dependence in one
polarity $R^{(+)}=V^{(+)}/I = R_{PC} + V_{temf}/I$ will differ
from the other one $R^{(-)}=V^{(-)}/I=R_{PC}-V_{temf}/I$ and will
cause the asymmetry of current-voltage characteristics. This
asymmetry can be more clearly observed on the current-voltage
characteristics derivative, for example differential resistance
$R_{D}=dV/dI$. Here, by analogy, we shall write
$R_{D}^{(+)}-R_{D}^{(-)}=2dV_{temf}/dI = 2d/dI
\int_{T_{0}}^{T_{PC}}(S_{1}-S_{2})dT\simeq
2R_{PC}d/dV\int_{T_{0}}^{T_{PC}}(S_{1}-S_{2})dT$. As mentioned
above, in the thermal regime there exists a linear relation
$k_BT=eV/2L^{1/2}$ between the contact temperature (at $T_0 \ll
T_{PC}$) and voltage. That is why the above expression can be
simplified after substituting the differentiation variable V by T
as
\begin{equation}
\frac{R_{D}^{(+)}-R_{D}^{(-)}}{R_{PC}}=\frac{e}{k_B\sqrt{L}}
[S_{1}(T_{PC})-S_{2}(T_{PC})]
\label{asym}
\end{equation}

Thus, the asymmetry of differential resistance of heterocontacts
in the thermal regime is proportional to the Seebeck coefficients
difference of the contacting metals. A more rigorous theory which
appeared later on confirmed this qualitative analysis.

The current-voltage characteristics asymmetry of heterocontacts is
especially clearly defined in substances with short mfp and large
Seebeck coefficients. These systems include heavy fermion
compounds, Kondo lattices, intermediate-valence systems. The
measurements of current-voltage characteristics asymmetry in such
conductors may give information of thermoemf behavior therein. For
this purpose one should use the second electrode of normal metal
whose Seebeck coefficients are considerably smaller than in the
above compounds, i.e. $S_1 \gg S_2$. Hence, the $R_D$ asymmetry
will be directly proportional to the Seebeck coefficient (see
Eq.\,(\ref{asym})) in the substance under study. This is of not
only the academic interest, for it can be used for analyzing the
$S$ behavior in the sample, especially when studying substances
available in small quantities for which the traditional methods of
determining the Seebeck coefficient are hard to apply.

We have considered the cases of contact heating under the action
of transport current. Naturally, this does not exhaust all thermal
effects which are interesting when investigating PCs. Another
important trend is the study of heat conductivity in shorts when
instead of (or together with) the potential difference a
temperature gradient is applied to the contact. Theoretical
treatment of heat transfer through PC in the ballistic regime
brought about a rather unexpected result. The second derivative of
the heat flux versus voltage proved to be directly related to the
electron-phonon interaction function. It is not that simple to
measure this derivative and it has not yet been done
experimentally.

However, some studies confirm the specific features of heat
conductivity in PCs. One of the contacts making electrodes was
attached to a heat exchanger while the other was held in vacuum.
Both electrodes carried miniature pickups to check the electrode
temperatures depending on the contact voltage. The temperature was
found to rise slightly with voltage in each electrodes, this
temperature being different for various polarities. When the
temperature difference versus voltage curve was plotted, it was
close in shape to the electron-phonon interaction function of the
contacting metals. Therefore, the above experiments evidence the
electron-phonon interaction effect on the heat conductivity of PCs
and confirms, though not directly, some theoretical conclusions.
It should be noted that the measured temperature difference has a
finer structure that can be related to the peculiarities of phonon
scattering and enables their investigation.

Now let us see what happens if a contact is affected only by the
temperature gradient. Let one electrode be held at helium
temperature $T_{0}$ and the other at $T$, the inequality $T_0<T$
being held. In this case the heat will be transferred from the
heated part through a constriction to the colder part. In other
words, phonons will propagate from the hot place where their
number is large to the cold place where their number is smaller.
The electron diffusion will proceed in the same direction, thus
causing the emf. However, when the contact is formed by electrodes
of the same metal, as stated previously, the potentiometer
connected to the ends of electrodes should read zero voltage.
Nevertheless, an experiment made according to this scheme
demonstrated that this not quite the case. Even if the contact is
homogeneous and the electrodes are maintained at different
temperatures, a potential difference will develop between their
ends. This occurs because in this case there is a circuit of
homogeneous in composition but strongly geometrically
inhomogeneous materials due to PC where kinetic processes are
specific. Thus, for example, the flow of phonons directed from the
hot part to the cold one drags" the electrons. This causes thermal
emf of the so-called phonon drag. It is "hard" for the phonons to
drag the electrons through the bottleneck of PC and the drag
thermoelectromotive force is suppressed \footnote{ As shown by
theoretical treatment, the diffusion part of thermal emf will be
also suppressed in PC, but less than associated with drag.}. In
the case concerned total thermoemf will be written as a difference
between the bulk and contact contribution. The latter is strongly
suppressed in the contact, so only the bulk contribution remains.
Hence, the potentiometer connected to the sample leads will read
the voltage to a higher extent governed by the drag thermal emf.
This offers probably a unique possibility of measuring the
absolute coefficient of thermal emf in a material-homogeneous
electric circuit.

\section{Electrical fluctuations in PCs}

Noises present an important problem in science and technology
since they dictate the sensitivity of existing radio-electronic
devices which, in its turn, limits the magnitude and accuracy of
measurements, the quality of processing various signals. The
investigation of sources and nature of noise makes it possible not
only to measure it, but also to suppress it, thus increasing the
signal-to-noise ratio for particular measurements or devices.

What is noise? It is an unwanted voltage or current of fluctuation
nature generated by the electronic device or some circuit
component. If this voltage is applied to an audio-frequency
amplifier, this voltage gives rise to a hiss in the loudspeaker.
Hence the name "noise" which is applied to such fluctuation
processes, though in most cases we do not actually hear it.

What are the current or voltage fluctuations due to? First of all,
the current carriers in a conductor are in constant motion. As a
result of it, fluctuation emf appears at the conductor ends. At
each particular moment of time this emf will have a definite
value, but averaged over a longer time it will be equal to zero.
This noise was called thermal or Nyquist noise, while the average
square emf $\langle V_n^2 \rangle$ arising at the resistance $R$
and temperature $T$ in the frequency range $\Delta f$ can be
written as $4k_BTR \Delta f$. The other type of noise is connected
with the discreteness of current carrier charge. Thus, for
example, the events of electron flying out of the cathode
(emitter) of electronic tube (transistor) form a sequence of
independent events, occur at random times and lead to the
so-called shot noise. At this time $\langle V_{n}^{2} \rangle
\propto eIR/C$, $e$ being the electron charge, $C$ the system
capacity, i.e. the shot noise is proportional to the current.

In an impurity-free semiconductor electrons and holes can appear
and disappear at random due to their generation and recombination.
This also causes fluctuation emf in the sample called
generation-recombination noise.

One more source of noise attracted attention first in the electron
vacuum tubes. It was attributed to the average emission rate
alteration causing microscopic current fluctuations. The emission
rate alteration can be treated as some effect of collective
interaction of electrons, as a result of which such oscillations
or flicker of current occur. This noise is known as flicker noise.
It is also characterized by the fact that the spectral power of
flicker noise density $S_{V}=\langle V_{n}^{2} \rangle/\Delta f$
is frequency dependent close to the $1/f$ law, that is why it is
also called $1/f$ noise. The nature of this noise is not quite
clear to date. Different models are suggested, particularly
recently the flicker noise in solids has been explained in terms
of two-level system which serve as elementary disorder carriers in
the crystal lattice.

The noise in electric contacts was studied rather in detail. In
1969 Hooge obtained an empirical formula for the spectral density
of contact noise at the frequency $f$: $S_{V}=\alpha V^{2}/Nf$
where $\alpha$ is a constant, $N$ is the total number of charge
carriers in the contact. The latter value can be found from the
formula $N=nv$ where $v$ is the constriction volume, $n$ is the
carriers concentration.  $v$ can be easily estimate from the
resistance $v\simeq d^{3}=(\rho/R)^{3}$ using the Maxwell formula
mentioned above. As a result we will obtain a more convenient
expression for $S_{V}$, viz.
 \[S_{V}\simeq \frac{R^{3}V^{2}}{n \rho^{3} f}\].

The spectral densities of noise have been also measured for PCs.
Thus, for the thermal regime of current flow it was show that
$S_{V}\sim V^{t}$ where $t\simeq 2$ at contact voltages up to a
few millivolts, whereas at higher voltages $S_V$  exponentially
increases. The latter fact is attributed to highly nonequilibrium
state of electrons and phonons in the vicinity of the contact when
the characteristic phonon energies are reached. The $S_{V}$
dependence on the sample resistance also appeared close to
$R^{3}$. Consequently, the noise behavior in PCs under thermal
regime has no peculiarities compared to those already known.

\begin{figure}[t]
\includegraphics[width=7cm,angle=0]{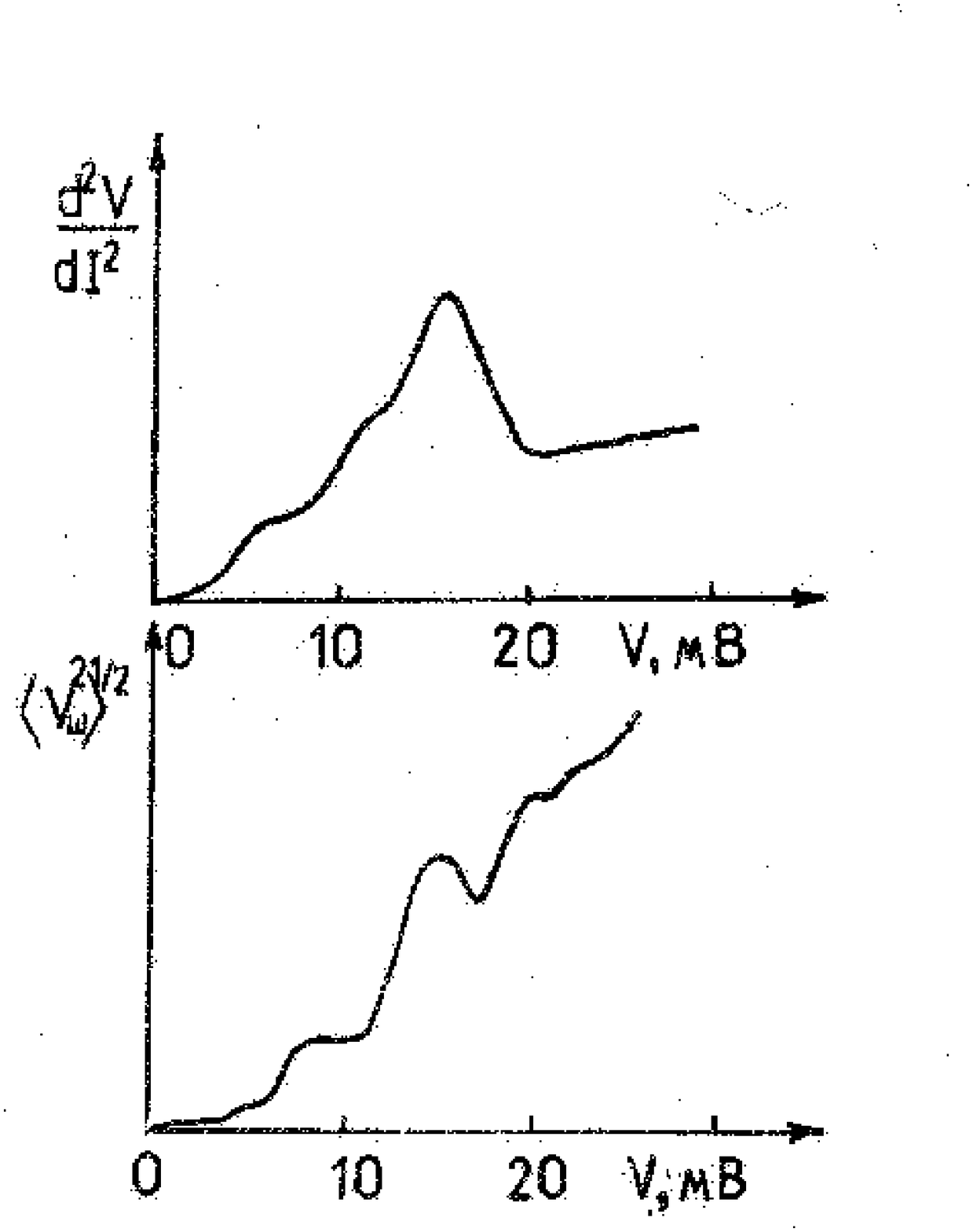}
\vspace{-1cm}
 \caption[]{\small Above: point-contact spectrum of a tin contact recorded
at the temperature 4.2 K.  Below: "noise" spectrum for the given
contact. In addition to  the monotonic growth of noise, the curve
display a number  of maxima and minima attributed to the
peculiarities of electron-phonon interaction in the contact.}
\label{fig14}
\end{figure}
\normalsize

 However, the study of $S_{V}$ dependence for contacts in the
ballistic regime yielded some new unusual information. It turned
out that in PC with $l_{e},l_{i}\gg d$ the spectral density of
noise does not increase monotonously but consists of a number of
extrema, both maxima and minima (Fig.\,\ref{fig14}). In this case
the minima positions in $\langle V_{n}^{2}\rangle$ at
$eV>\varepsilon_{D}$ corresponded to the energy of multiphonon
scattering of electrons. It was suggested that such multiphonon
generation in the constriction corresponds to the establishment of
some coherent (ordered) regime of phonon emission at the
realization of which the noise decreases. For the energies
$eV<\varepsilon_{D}$ the $\langle V_{n}^{2} \rangle$ also has some
minima attributed to the one-phonon normal electron scattering in
the coherent regime. It turned out that the positions of minima
correspond to the energy at which the phonon dispersion curves
cross the Brillouin zone boundary, i.e. the wave vector of phonons
is maximum and their velocity is minimum. The latter circumstance
promotes the accumulation of phonons in the vicinity of contact
and the establishment of coherent regime of emission. On the other
hand, the positions of maxima in $\langle V_{n}^{2} \rangle$ at $
eV<\varepsilon_{D}$ coincide with the minimum energy of phonons
for which the scattering with Umklapp is possible. Such scattering
correspond to the Bragg reflection of the electron wave from the
atomic plane in the lattice. At this time the electron momentum
changes by the inverse lattice vector $2 \pi \hbar/a$, i.e. equal
to the double maximum momentum of phonon. The above processes are
non-correlated because not only phonons but the lattice as a whole
participate in the electron scattering events and the noise
increases. The observed reproducibility of extrema position on the
curve $\langle V_{n}^{2} \rangle$ for contacts of different
resistance and their relation to the characteristic phonon
energies more precisely to the singular points of the phonon
dispersion curves, makes it possible to speak about the noise
spectroscopy of separate groups of phonons in metals.

\section{PCs in a high-frequency electromagnetic field}

 For a considerably long time PCs have been used in
microwave engineering  as detectors, mixers and harmonic signal
sources. Such contacts are formed by means of thin wire pressed
against a smooth metal surface. The wire serves as a kind of
antenna through which the high-frequency current penetrates in the
bulk of the contact. As is already known, the current-voltage
characteristics of PCs deviate from the Ohm law both in the
ballistic and thermal regimes. This is why the contact is a
nonlinear element and possesses the above detecting and other
properties. However, the physics of processes proceeding in the
contact under the action of microwave irradiation is of
considerable interest.

 The first thing done by researchers was to analyze the form of
current-voltage characteristics derivatives at high frequencies.
The experimental technique in this case differs a little from that
described in Chapter 8 and used for audio frequencies. As is seen
from Eq.\,(\ref{d2v}), in addition to the signal proportional to
$d^{2}V/dI^{2}$ at the frequency $2\omega$, there is a dc signal
of the second derivative. To detect this constant component of
$d^{2}V/dI^{2}$, the microwave signal is modulated by means of
some chopper. For example, use is often made of a rotating disc
with holes to pass radiation. Then, using a circuit similar to
that described in Chapter 8, the signal proportional to
$d^{2}V/dI^{2}$ is separated at the chopping frequency.

 Naturally, the PC resistance is much lower then that
of the surrounding medium, so when microwave field is applied, a
high-frequency current will appear in the constriction and will
play the role of modulation signal. Therefore, for a contact in
the ballistic regime one can measure the second derivative
proportional to the electron-phonon interaction function, but only
at high frequencies. Such experiments were carried out and showed
that the $d^{2}V/dI^{2}$ obtained in this way was identical to the
second derivative recorded at low frequency. Distinctions appeared
when the frequency became as high as giga- and tera hertz or the
period of oscillation reached $10^{-9}-10^{-12}$s. Here
$d^{2}V/dI^{2}$ began to change: first the background level
decreased, then, at still higher frequencies, the spectral
features began to broaden. These change in $d^{2}V/dI^{2}$ curve
can be understood if we compare the microwave signal period with
times of quasi-particles relaxation in the contact. Thus, the
electron-phonon scattering occurs within $10^{-13}$ to $10^{-14}$
s which can be readily obtained dividing the mfp the Fermi
velocity of electron. The phonon-electron scattering processes are
as many times slower as sound (phonon) velocity is smaller than
the Fermi one and are characterized by times $10^{-9}$ to
$10^{-10}$ s. They also affect the point-contact spectrum causing
the presence of phonon component background therein. Consequently,
when the irradiation frequency approaches $10^{9}$ Hz that is
comparable with the frequency of phonon-electron relaxation, the
phonon reabsorption processes will slow down because they will not
be able to trace rapid oscillations of the electron gas under the
action of emf induced by external radiation. As a result the
background level in the $d^{2}V/dI^{2}$ curve recorded at
microwave frequency should decrease. With further frequency
increase to the level comparable with the frequency of
electron-phonon interaction (in this case the energy quantum of
radiation is comparable to the phonon energies) the processes in
the contact should be considered  in terms of corpuscular theory
as the absorbtion and emission of phonons under the action of
radiation quanta. As a result of it further increase of frequency
will cause smearing of $d^{2}V/dI^{2}$ phonon spectral features
and the appearance of satellites at higher frequencies.

When passing to the thermal regime, the nonlinearity of
point-contact conduction is connected with the effect of heating.
In this case the characteristic time is determined by thermal
relaxation  processes in the contact. It depends on the contact
size $d$, specific heat capacity $c$, heat conduction of material
$\lambda$ and can be evaluated by the formula
$\tau=cd^{2}/\lambda$. At characteristic size of contact $10^{-8}$
m, heat capacity and heat conductivity of metals the value of
$\tau$ is about $10^{-9}$ s. Therefore, in the frequency range
$10^{9}$ Hz and higher the $d^{2}V/dI^{2}$ curve should change its
form as compared with that obtained at low frequencies. This is
caused by the fact that at the frequencies above those of thermal
relaxation the contact temperature no longer traces the modulating
high-frequency current. As a result the possibilities of
temperatures modulation spectroscopy described in Chapter 13 are
limited by the above frequency.

Thus, the study of contact conductivity at high frequencies makes
it possible to clear out the nature of various non-stationary
processes in metals, to determine the time of phonon-electron
collision, to elucidate the quantum limit effect in detecting for
the ballistic regime. In the thermal regime it enable the
investigation of thermal relaxation processes in small-size
conductor.

\section{Study of PCs on semimetals and semiconductors}

The distinctive features of substances relating to semimetal and
semiconductor is, first of all, low concentration of charge
carriers as compared with common metal. Thus, in typical
semimetals, bismuth, antimony and arsenic, it changes from
$3\times10^{17}$ (Bi) to $2\times10^{20}$ (As) per 1 cm$^{3}$,
whereas in metals it is $10^{22}-10^{23}$. In semiconductors these
value can be still lower by several orders of magnitudes.
Accordingly, the Fermi energy decreases from several
electron-volts in metals to millielectron-volts in semimetals.
Secondly, the effective mass of charge carriers in semimetals, due
to the peculiarities of their band structure, can be more then an
order of magnitude lower then its value in common conductors. The
latter, together with low Fermi energy, causes the increase of the
de Broglie wavelength in semimetals up to tens of nanometers, i.e.
it becomes comparable with the contact dimensions. Thirdly, small
effective mass leads to the fact that the Larmor radius of
electron trajectory decreases to hundreds of nanometers already in
an easily attainable magnetic field about 1 Tesla. Thus, an
interesting situation is created in semimetals when all the above
lengths become comparable with the constriction dimension. This
leads to the influence of quantum interference effects on the
conductivity of contact since due to a large wavelength the
quasi-particles can no longer be considered as point ones and
their wave properties should be also taken into account. Besides,
the magnetic field effect on the charge carrier trajectories in
the vicinity of the contact should be appreciable, which will also
affect the characteristics being measured. Thus, for example,
under the field effect we observed transformation of the
electron-phonon interaction spectrum in antimony: the change of
intensity, half-width and energy position of maxima depending on
the relation between the Larmor radius of carriers and their mfp
and contact size.

 On the other hand, a number of antimony contact displayed the
change of spectrum sign, i.e., $d^{2}V/dI^{2}$ became negative but
the extrema due to the electron-phonon interaction were retained.
Such contacts were distinguished by small elastic mfp which was of
the same order as the de Broglie wavelength. As a result, due to
the quantum interference of electron waves, the current carriers
are localized and the system conductivity decreases. The
electron-phonon interaction processes destroy localization and
increase conductivity which corresponds to the decrease of
differential resistance and, consequently, to the negative sign of
$d^{2}V/dI^{2}$. This mechanism is most effective at the energies
with many phonons, i.e. at the phonon spectrum maxima. Finally,
under the localization regime for the above contacts the recorded
spectrum shape will be close to the common one as if turned upside
down. As stated above, this interesting effect was observed for
contacts made of antimony which provide to be sufficiently
convenient and workable material for making shorts.

Let us dwell on the point-contact study of semiconductors where so
far only the first steps are made. Note that this problem is much
greater tackled theoretically than experimentally where the
investigation can be counted on the fingers of one hand. The
matter is that the above technology of making PCs between metals
is not well suitable for semiconductors. This is due to the much
stronger sensitivity of the surface properties of semiconductors
to deformation, pressure, imperfect structure, impurities which
are always present to some extent when the contact is formed
mechanically. Even a perfect surface is a "defect" relative to the
bulk material because the boundary atoms on the one hand interact
with the like ones, and on the other hand interact with the atoms
of the surrounding medium. In this way the surface state appear on
the atomically pure surface of the semiconductor whose energy
levels lie in the forbidden zone (Tamm or Schottky levels). In
real crystals, along with the above levels, the additional ones
appear due to the impurities, defects, adsorbed atoms, etc. The
density of such levels depends on surface boundary and can vary in
wide range. The presence of surface energy levels result in the
formation of near-surface layer of the space charge whose
thickness is determined by the screening distance $r_{S}$
\footnote{In conductor, the electron gas screens separate charges
whose radius of action is in this case limited by the distance
$r_{S}$ depending on electron concentration as $r_{S} \sim
n^{-1/2}$. In metals the screening distance is only a few decimal
fractions of \AA, while in semiconductors it can reach tens and
hundreds of \AA.} depending on the carriers concentration. At this
time layer with the main carriers concentration other than in the
bulk appears on the surface. As a result of it, when mechanically
contacting two semiconductors, the point-contact properties will
be to large extend determined by the surface layer properties and
differ from the bulk ones. To avoid it, the lattices should
perfectly coincide during the mechanical contact which hardly
probable. For this reason the characteristics of semiconductor PCs
strongly vary from contact to contact and are poorly reproducible,
thus impeding the studies. Nevertheless, attempts to advance in
this direction are continued. Thus, technique of creating
micro-constrictions has been implemented in semiconductor
heterocontacts. Two-dimensional micro-constrictions less then one
micron in size have been obtained by electron-beam lithography.

\section{Spectroscopy of superconductor}

 So far we considered the current-voltage characteristics of
contacts formed by normal metals. However, at low temperatures
many metals and alloys become superconducting. It should be noted
that researchers began to study contacts between two
superconductors long before the advent of point-contact
spectroscopy. To tell the truth, the scientists were more
interested in the properties of superconducting condensate formed
by Cooper pairs. To break pair, it should be given some energy
equal to or greater then the energy gap \footnote{In common
superconductors the electrons interact via phonon and form the
so-called Cooper pairs which ensure a non-dissipative charge
transfer. As the pairs are Bose-particles with the spin  equal to
zero, they are all "condensed" at Fermi level. The   energy
required for breaking a pair is equal to the energy gap  arising
in the electron density of states in superconductor at  the Fermi
level. The energy gap value, as well as those of   critical
temperature and other parameters, is one of the most important
superconductor characteristics.} in the electron density of states
equal not more than several millielectron-volts. The
current-voltage characteristics of superconducting contacts first
of all reveal the features connected with the presence of the
energy gap. Such features were studied by many experiments in all
simple superconducting metals.

 A question arises whether the phonon features will show up in the
second derivatives of current-voltage characteristics of
superconducting contacts, i.e. whether the point-contact
spectroscopy of superconducting state is possible. The answer
proved to be positive. For simple superconducting metals (Pb, Sn)
it was demonstrated that the point-contact spectra taken bellow
the temperature of superconducting transition $T_{c}$ displayed
the features associated with the electron-phonon interaction. At
this time the form of the spectrum practically coincide with
similar dependence for the same contacts in the normal state. Only
at the energies low compered with the Debye one there appear sharp
features due to the energy gap manifestation. But they do not
greatly distort the electron-phonon interaction spectrum since the
phonon are located at higher energies.

 At the first sight it seems that point-contact spectroscopy in
the superconducting state yield no new information. However, not
all the metals, especially alloys and compounds with high critical
parameters: superconducting transition temperature, critical
magnetic field, can be studied at low temperatures in the normal
state. Therefore, the principal possibility of the point-contact
spectroscopy application for investigation the superconductors
considerably extends its scope of use to the study of the
high-temperature superconductors.

 It is noteworthy that the point-contact spectroscopy of
superconductors has offered new opportunities. It tuned out that
in some metals and alloys whose point-contact spectra in the
normal state do not display any spectral features due to the short
electron mfp, in the superconducting state there appear peaks
arranged in the region of characteristics phonon frequencies. This
was found out experimentally when measuring contacts of Nb, Ta,
Nb$_{3}$Sn, NbSe$_{2}$. The nature of these features is somewhat
different from that in the ballistic contact. They are connected
with the influence of nonequilibrium phonons generated by
electrons on the superconducting properties of metal near the
constriction. The phonons with small group velocities more slowly
emerge from the contact vicinity, cause local overheating and
decrease of the energy gap. Therefore, at these energies features
will appear in the current-voltage characteristics derivative.
This makes it possible to carry out the spectroscopy of "slow"
phonons which correspond to the flattening of dispersion curves
where $d\varepsilon/dq$ tends to zero (see Fig. 2a).

\section{New trends of research}

 In the preceding chapters we discussed the possibilities of
point-contact spectroscopy when studying mainly simple metals and
some alloys. It was shown that this technique makes it possible to
obtain detailed (spectroscopic) information on the mechanisms of
electron scattering in conductor. At the same time the recent
advances of solid-state physics are associated with the synthesis
and study of new classes of compounds. These include the
heave-fermion and mixed-valence systems, so-called Kondo lattices,
new high-temperature superconductors and many others. Naturally,
the most promising trends in point-contact spectroscopy are also
connected with their study.

The substances in question are noted for the fact that the
conventional band scheme proceeding from the concept of free
electrons in the conduction band and the localized nearly atomic
states of lower-lying shells is not suitable for their
description. In these substances the inner shells near the Fermi
level lose their stability and their electrons acquire a partially
band character. As a result the electron density of states
strongly increases and can have a sharp maximum near the Fermi
level. In a number of case the presence of such narrow maximum in
Kondo lattices is attributed of many-particle Abrikosov-Suhl
resonance caused by the interaction between localized f- or
d-electrons and conducting electrons. A high increase of the
electron density of states corresponds to the appearance of
electrons with a larger effective mass. In simple terms, these
conduction electrons are not fully "untied" from the atoms and
their field-effected motion is more hindered. The presence of such
electrons stipulates the unusual properties of these materials.

Theoretical treatment of the kinetic properties of such substances
is based on taking into account their electronic spectrum
peculiarities near the Fermi level. It is impeded by the absence
of direct experimental data on the electron density of states in a
narrow layer on the Fermi surface. The scientists were tempted by
the idea to use PC for recording the energy spectra not only of
phonons but also of electrons.

A number of research groups abroad are investigating the compounds
with heavy fermions and intermediate valence in heterocontact with
normal metal. It was discovered that for the substances with a
maximum electron density of states $N(\varepsilon)$  near the
Fermi energy $\varepsilon_{F}$ the point-contact differential
resistance $R_{D}(V)$ curve has deep minimum at $V=0$. At the same
time, a sharp maximum of $R_{D}$ at $V=0$ appear in the substances
with a low density of states or with gap, i.e. $N(\varepsilon)=0$
at $\varepsilon_{F}-\Delta<\varepsilon<\varepsilon_{F}+\Delta$
where $2\Delta$ is the gap width. It also appeared that the widths
of $R_{D}(V)$ maxima or minima is close to the assumed widths of
$N(\varepsilon)$ features. This suggested the possibility of
registering the $N(\varepsilon)$ features near the Fermi level in
such substances using the point-contact spectroscopy. However, not
everything is so simple in this problem. Firstly, the electron mfp
in the systems under consideration are small enough and one should
take into account the thermal effects occurring in PCs. Secondly,
there is not microscopic theory that allows for the effect of
electronic spectrum peculiarities on the point-contact kinetic
phenomena. The experiments carried out by other groups showed,
however, that the thermal regime is realized in the contacts of
such materials and that the main nonlinearity of the
current-voltage characteristics is determined by the
temperature-depended part of substance resistance. Thus, for
example, the $R_{D}(V)$ behavior for contacts made of number of
heavy-fermion compounds, such us CeCu$_{2}$Si$_{2}$, UPt$_{3}$,
UBe$_{13}$ and others, precisely duplicates the $\rho(T)$
dependence in bulk material. Thus, the relation of $V$ and $T$ in
a contact, typical on the thermal regime, is evident. Besides, it
was shown that one should be very careful when treating the result
obtained on heterocontacts \footnote{This experimental geometry is
often used since the above  compounds are frequently rather small
samples and cannot be  adequately shaped, so it is simple to make
a needle of another material.} between a normal metal and specific
substances, for here one should also successively take into
account the thermoelectrical effects considered in Chapter 13.

 There is probably only one convincing example of how the authors
managed to obtain well-defined spectral features in the
point-contact spectra of intermediate-valence compound CeNi$_{5}$.
Such substances contain rare-earth atoms with partially filled 4f
shell and +3 valence. Since the ion states with filled or empty
shell are most stable, the state Ce$^{3+}$ will compete with the
state Ce$^{4+}$ when the f-shell in cerium is empty. Thus, the
Ce$^{3+}$ ions in the compound CeNi$_{5}$ can also exist in the
Ce$^{4+}$ state, giving out the excess electron. As a result,
there is constant competition between these two state, Ce$^{3+}$
and Ce$^{4+}$, so the valence of Ce ions is between 3 and 4 is
called intermediate. There exists a model according to which the
energy levels of the Ce$^{3+}$ and Ce$^{4+}$ states are arranged
near the Fermi level but on the opposite sides, thus increasing
the density of state at the corresponding energies. The electrons
take part in the transitions between these states or levels with
difference valence. These transitions affect the conductivity of
substance and should be observed on the point-contact spectra.
Actually, among with the phonon features, the $d^{2}V/dI^{2}$
curves of CeNi$_{5}$ contacts reveal small maxima in the low
energy region. These maxima can be related with scattering by
different-valence configuration of the Ce ion. In this case the
$d^{2}V/dI^{2}$ curve of CeNi$_{5}$-normal metal heterocontact
displays a large asymmetry of features in question depending on
the voltage polarity. In terms of the model under consideration,
this is attributed to the resonance scattering by
different-valence state of the Ce ion with allowance for the
specific form of the electron distribution function in PC (see
Fig.\,\ref{fig5}b).

 On the other hand, there is an alternative approach to the
explanation of mentioned features observed in the PC spectra of
CeNi$_{5}$. This approach is connected with taking into account
strong spin correlations in this compound. Their important role
was indicated by the low temperature behavior of CeNi$_{5}$
resistance proportional to $T^{2}$ attributed to the scattering by
spin fluctuation or paramagnons. The latter should lead to the
linear dependence of the initial $d^{2}V/dI^{2}$ section that was
observed on all curves in the absence of negative "zero-bias"
anomaly, which, as noted in Chapter 7, indicates low quality of
contacts.

 As to the $d^{2}V/dI^{2}$ asymmetry in heterocontacts based on
CeNi$_{5}$, the detailed study on single crystals has demonstrated
its greater relation to the thermoelectric effects. This evidence
by the fact that the asymmetry corresponds to the thermoemf sign
in CeNi$_{5}$ and its value for the two principal directions
correlates with that of the Seebeck coefficient. Thus, further
investigation of CeNi$_{5}$ by means of PC will enable better
understanding of these interesting systems.

 Nevertheless, it should be noted that the discussed possibility
of refining the band structure of the compounds under
consideration by means of point-contact spectroscopy is of
interest due to a high resolution about 1 meV. The traditional
X-ray or photoelectron techniques used for the study of band
structure operate in significant larger energy scales with
resolution not higher than hundreds of millielectron-volts.

\begin{figure}[t]
\includegraphics[width=8cm,angle=0]{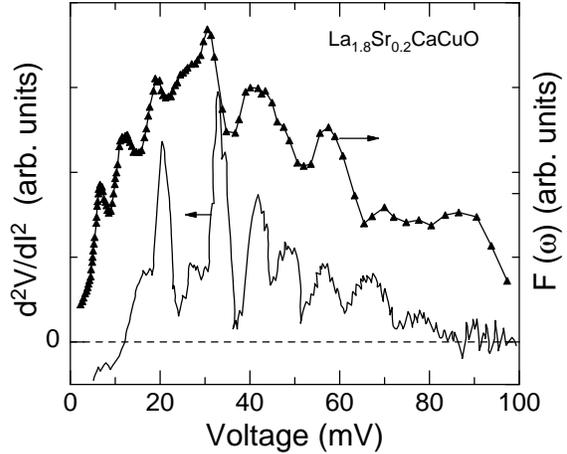}
\caption[]{\small Point-contact spectrum of high-temperature
superconductor La$_{1.8}$Sr$_{0.2}$CuO$_{4}$ (solid curve). A
number of narrow spectral lines and the spectrum boundary near 80
mV are well defined. The initial spectral part up to 10 mV contain
powerful bursts due to the manifestation of the superconductor
energy gap not shown in the figure. Neutron phonon density of
states (symbos) is shown for comparison.} \label{fig15}
\end{figure}
\normalsize

The point-contact technique did not stand of the events associated
with the discovery of high-temperature superconductors. One of the
first compounds of this kind, La$_{1.8}$Sr$_{0.2}$CuO$_{4}$ was
studied by means of point-contact spectroscopy. It was rather
simple to obtain the point-contact spectrum of phonons
(Fig.\,\ref{fig15}). It was found that it extends up to the
energies of about 80 meV and comprises six sharp peaks
corresponding to the phonons with the small group velocities. The
electron-phonon interaction estimated by the relative differential
resistance increase of PC when the energy rises to
$\varepsilon_{D}$ has shown this interaction to be almost the
order of magnitude stronger than in common superconductors.
Consequently, the superconductivity in the compound in question is
more than likely due to the electron-phonon interaction.  On the
other hand, the analysis of current-voltage characteristics of
contact from such superconductors below the critical temperature
enables the determination of their energy gap. The information
obtained thereat relates to the substance region with the size
about the contact diameter directly adjacent to the constriction.
This makes it possible to investigate the properties of separate
superconducting clusters whose sizes measure only hundreds \AA.
The knowledge of the contact dimensions enable us to estimate the
critical current densities. It turned out that this density
reaches $10^{8} $A/cm$^{2}$ in the superconducting clusters
investigated by the point-contact spectroscopy, whereas in the
bulk sample the critical current density does not exceed $10^{3}
$A/cm$^{2}$. Apparently, the latter fact points to the presence of
weak contacts between superconducting granules in such systems.
Therefore, PC can be also used as an efficient probe which enables
the study of cluster and other polyphase compounds.

 It is necessary to note the even growing interest to the study of
heterocontacts including those between pure metals. It is clear
that the point-contact spectrum of heterocontact between different
metals should contain the peculiarities due to one and another
electrodes, i.e. as a result the $d^{2}V/dI^{2}$ dependence can be
presented as a sum of spectrum from each materials. Such
experiments were made soon after obtaining first point-contact
spectra. The researchers observed the change the partial
contribution into the spectrum from one or another substance for
various contacts with different resistances. This was attributed
to the fact that the point-contact vicinity can be at random
mostly filled with one or another metal. However, theoretical
works in this field have demonstrated that the heterocontacts have
their own physics. Firstly the intensities of one or another metal
spectra in the heterocontacts are inversely proportional to the
Fermi velocity, i.e. in other words, to the time of electrons
transit or interaction in the contact. Secondly, the effects of
electron trajectories reflection or refraction appear in the
heterocontact (due to the different electron momenta) as is the
case when the light passes through the boundary between two media.
As a result, as shown by the estimates, it can lead to a
considerable change in the metal spectrum both in intensity and
shape. The detailed comparison of theoretical calculations with
experiments on heterocontacts yields information on the electron
scattering at the metals interface.

Besides, the heterocontacts develop a whole number of other
interesting effects connected with the observation of spectrum
asymmetry \footnote{ In this case the asymmetry is not connected
with the thermal  emf since the contacts are in the ballistic
regime.} depending on the applied voltage polarity. One of such
effects detected recently is related to the electron drag in the
contact by the nonequilibrium phonons which enable the study of
their relaxation processes with the aid of PCs.

\section{Conclusion}

(Note, it was written 15 years ago.)

 The above examples explain the achievements and prospects of the
development of point-contact spectroscopy for the study of the
electron and phonon system kinetics which determines the main
electrical properties of conductors. This technique is becoming a
powerful and efficient tool for the study of highly nonequilibrium
phenomena in conductors, especially those with limited space
dimensions. The latter is of interest for modern microelectronics
where the dimensions of separate components are already below the
micron range.

 Note that the point-contact measurements are also helpful in all
the case when one deals with constrictions of different types.
Thus, they enable obtaining information on the state of metal in
the regions of current concentrations, mean free paths, contact
dimensions, etc., i.e. to have some "rated" characteristics of
contact, the knowledge of what researcher or engineer is dealing
with.

The best evidence of the efficiency and importance of work
carried out in this field is the wide geography of point-contact
spectroscopy research. Thus, in the USSR both the experimental and
theoretical studies are extensively carried out at the Institute
for Low Temperature Physics and Engineering of the Ukrainian
Academy of Sciences, which takes the leading place in the world in
this respect. This is the birthplace of the point-contact
technique where further fruitful research is under way.

Experimental groups of "point-contact" physicists successfully
work in the Kurchatov Institute of Atomic Energy and the Institute
of Solid-State Physics of USSR Academy of Sciences. Their
principal interests is connected with transition metals and alloys
on their base. There are also groups of the theorists dealing with
the calculation of the point-contact spectra.

A number of important results have been obtained by Dutch
scientists from the University of Nijmegen who, in particular,
proposed the "needle-anvil" technique. Now this group has moved to
Grenoble, to the Laboratory of High Magnetic fields in the
Institute of Solid-State Physics of the Max-Planck Society. The
principal attention here is paid to the study of PCs on the base
of compounds with rare-earth ions and their behavior in high
magnetic fields. The researchers of the University of Cologne
(Germany) use PCs for the investigation of heavy-fermion
compounds, Kondo lattices and mixed-valence systems. The group
from the Higher Technical School (ETH) in Z\"urich (Swiss) works
in the same direction.

The physicists in the USA and Canada are actively working on the
theory of point-contact spectroscopy and calculation of
point-contact spectra of simple metals. Experimental study of PCs
from some semiconductors and pure rare-earth metals are carried
out in the Cavendish laboratory in Cambridge (England). At the
Institute of Experimental Physics in Kosice (Czecho-Slovakia)
scientists are successfully developing various trends in
point-contact spectroscopy in close collaboration with the
researchers from the Institute for Low Temperature Physics and
Engineering of the UkrSSR Academy of Sciences.
 Recently, the scientists of Japan joined in the point-contact
studies. The group in Tokyo and Sendai are engaged in the
point-contact spectroscopy of Kondo lattices. There are number of
other places where scientists are not only interested in these
technique but are trying to use it. Within the last 15 years of
its development, the point-contact spectroscopy, together with the
other spectroscopic techniques, has become inseparable part of
physical research. More than two hundreds papers on this subject
have been already published in the leading physical journals all
over the world. The papers devoted to point-contact spectroscopy
are presented at practically all conferences dealing with the
electronic properties of solid. It is hardly too much to say that
the point-contact spectroscopy method has been adopted by
physicists all over the world.

\section*{Literature}

Holm R., Electric contacts handbook. M., Foreign Literature
   Publishers, 1961 (in Russian).

Kaganov M.I., Electrons, phonons, magnons. M., Nauka, 1979 (in
   Russian).

Khotkevich A. V. and Yanson I. K.,  {\it Atlas of Point Contact
Spectra of Electron-Phonon Interaction in Metals}, Kluwer Academic
Publisher, Boston, 1995.

Kittel C.,  {\it Introduction to Solid State Physics} by John
Wiley \& Sons, Inc., 1971.

Verkin B.I., Point-contact spectroscopy of metals and alloys,
Priroda, No. 10,  1983 (in Russian).

\vspace{0.5cm} {\bf Topical reviews}~~(added) \vspace{0.5cm}

Duif A., Jansen A. G. M. and Wyder P.,  J. Phys.: Condens. Matter
{\bf 1 } 3157 (1989).

Jansen A. G. M., van Gelder A. P. and Wyder P.,  J. Phys. C {\bf
13} 6073 (1980).

Naidyuk Yu. G. and Yanson I. K.,  J. Phys.: Condens. Matter {\bf
10 } 8905 (1998).

Yanson I. K.,  Sov. J. Low Temp. Phys. {\bf 9} 343 (1983).

Yanson I. K.,  Sov. J. Low Temp. Phys. {\bf 17} 143 (1991).

Yanson I. K. and Shklyarevskii O. I.,  Sov. J. Low Temp. Phys.
{\bf 12} 509 (1986).

\begin{figure*}[]
 \begin{center}
\includegraphics[width=12.5cm,angle=0]{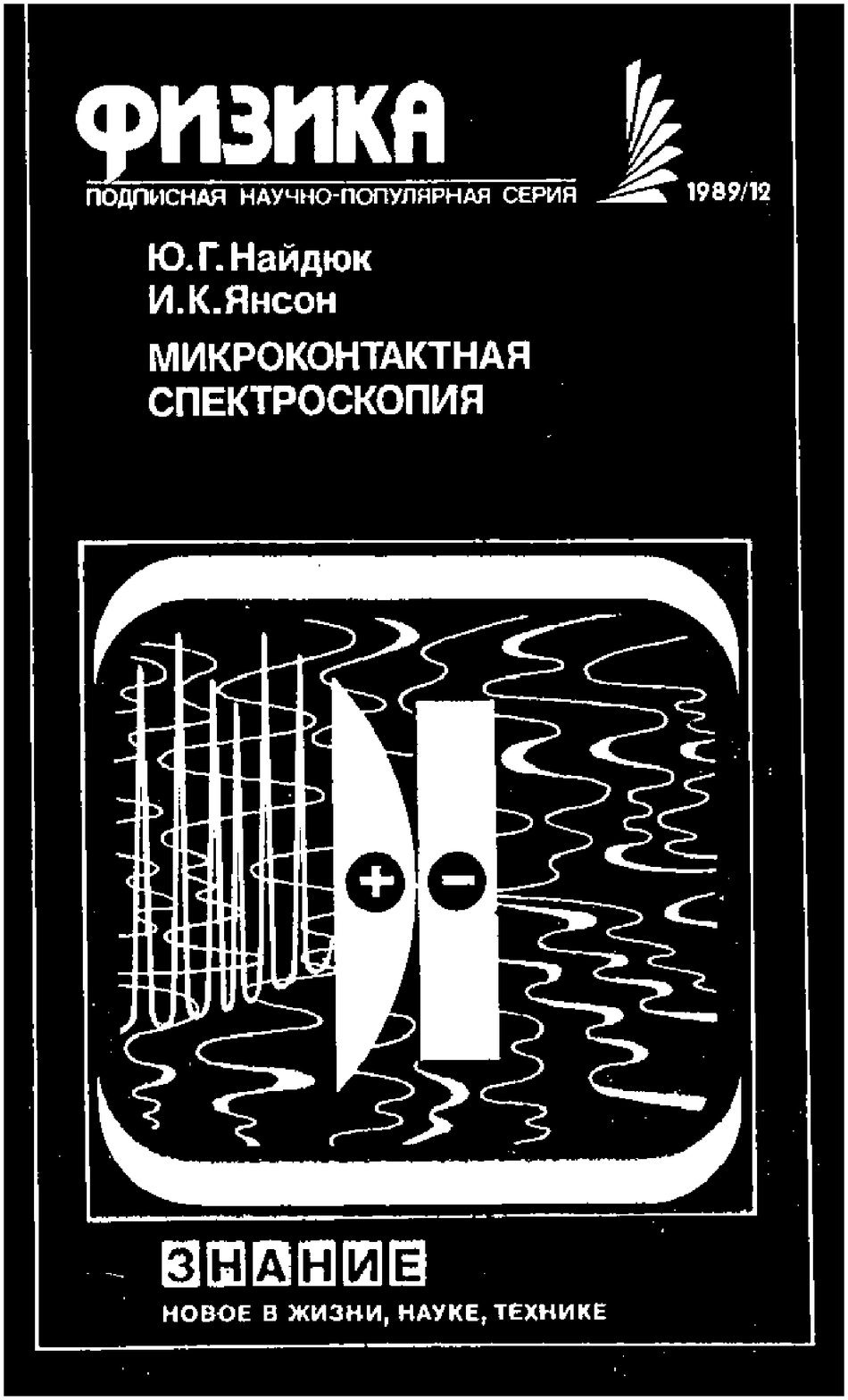}
\caption{Original brochure cover.}
 \end{center}
\end{figure*}

\end{document}